\documentclass[10pt,journal]{IEEEtran}
\IEEEoverridecommandlockouts

\usepackage[T1]{fontenc}
\usepackage{cite}
\usepackage{svg}
\usepackage{amsmath,amssymb,amsfonts}
\usepackage{bm}
\usepackage{graphicx}
\usepackage{textcomp}
\usepackage{xcolor}
\usepackage{makecell} 
\usepackage{multirow}
\usepackage{calrsfs} 
\usepackage{url} 
\usepackage{pifont}
\usepackage{tabularx}
\usepackage{booktabs}
\usepackage{algorithm}
\usepackage{algpseudocode}
\usepackage{amsthm}
\newtheorem{theorem}{Theorem}

\theoremstyle{remark}

\usepackage[breakable]{tcolorbox}   
\usepackage{capt-of}                
\usepackage{hyperref}  
\algrenewcommand\algorithmicrequire{\textbf{Input:}}
\algrenewcommand\algorithmicensure{\textbf{Output:}}
\algrenewcommand\algorithmicindent{1em}

\usepackage{tikz}

\def\BibTeX{{\rm B\kern-.05em{\sc i\kern-.025em b}\kern-.08em
    T\kern-.1667em\lower.7ex\hbox{E}\kern-.125emX}}

\usepackage{xcolor}
\usepackage{listings}
\usepackage{framed}

\definecolor{promptbg}{HTML}{F6F7F9}
\definecolor{promptframe}{HTML}{B9BEC6}
\definecolor{promptfield}{HTML}{7A2E8E}
\definecolor{prompttag}{HTML}{1B3A6B}
\definecolor{promptstr}{HTML}{1F6B3B}

\lstdefinestyle{trace}{%
  basicstyle=\ttfamily\fontsize{8}{9.3}\selectfont,
  frame=tb, framerule=0.7pt, rulecolor=\color{promptframe},
  xleftmargin=0.4em, xrightmargin=0.4em,
  framexleftmargin=0.4em, framexrightmargin=0.4em,
  numbers=none,
  columns=fullflexible, keepspaces=true, showstringspaces=false, upquote=true,
  breaklines=true, breakatwhitespace=true, breakindent=1em,
  aboveskip=0.4\baselineskip, belowskip=0.2\baselineskip,
  alsoletter={_},
  morekeywords={storm_active,malicious_drop_prob,tighten,lyapunov_V,lyapunov_W,
                queue_hold_threshold,attendance_reasoning,expected_attendance,
                event_time,reasoning,reserve_servers,queue_len},
  keywordstyle=\color{promptfield},
  morestring=[b]", stringstyle=\color{promptstr},
  moredelim=**[s][\color{prompttag}\bfseries]{[}{]},
  literate={->}{{\,$\rightarrow$\,}}3
}

\colorlet{shadecolor}{promptbg}
\newenvironment{tracebox}[1]{%
  \par\addvspace{0.6\baselineskip}\noindent
  {\footnotesize\bfseries #1}\par\nobreak\vspace{-0.55\baselineskip}%
  \snugshade
}{%
  \endsnugshade\par\addvspace{0.4\baselineskip}%
}

\newcommand{\cscarrow}{\textcolor{promptframe}{$\downarrow$}}
\newcommand{\csctier}[1]{\textbf{\footnotesize #1}}
\newcommand{\cscdir}[1]{{\ttfamily\footnotesize\color{promptfield}#1}}
\newenvironment{cascade}{%
  \footnotesize\setlength{\tabcolsep}{0pt}%
  \begin{tabular}{@{}r@{\hspace{7pt}}p{\dimexpr\linewidth-58pt\relax}@{}}%
}{\end{tabular}}

\begin{document}
\title{\huge From Reactive to Agentic Resilience: Multi-Timescale AI-RAN Orchestration Under Signaling Storms
}

\author{
\IEEEauthorblockN{
    Sotiris Chatzimiltis\IEEEauthorrefmark{1}, \textit{Graduate Student Member, IEEE}, Mahdi Boloursaz Mashhadi\IEEEauthorrefmark{1}, \textit{Senior Member, IEEE},\\
    Mohammad Shojafar\IEEEauthorrefmark{1}, \textit{Senior Member, IEEE},   
    and Rahim Tafazolli\IEEEauthorrefmark{1}, \textit{Fellow, IEEE}
}\\
\IEEEauthorblockA{
    \IEEEauthorrefmark{1}5G/6GIC, Institute for Communication Systems (ICS), University of Surrey, Guildford, UK \\
    \{sc02449, m.boloursazmashhadi, m.shojafar, r.tafazolli\}@surrey.ac.uk
}
}
\maketitle

\begin{abstract}
Next-generation open radio access network (Open RAN) and AI-RAN deployments turn the RAN into a disaggregated, compute-aware platform whose service continuity depends on many control loops acting at different timescales. Resilience in this setting is best seen as a lifecycle spanning anticipation, absorption, adaptation, recovery, and evolution. Yet, existing mechanisms remain stage-specific and are rarely coordinated across the control hierarchy. This paper argues that agentic AI offers an orchestration layer that coordinates this lifecycle, turning operator intent into coordinated action while keeping slow language-model reasoning off the fast control path. Using mass-attachment signaling storms as a running study, we build a discrete-event Open RAN attach simulator that captures retry amplification, shared-compute contention, and non-instantaneous provisioning, and we propose a three-tier framework in which an orchestrator at the service management and orchestration layer sets a standing policy, a per-site non-real-time agent judges storms and anticipates scheduled crowds, and a deterministic near-real-time loop enforces every decision. Across experiments, the agentic controller approaches the resilience of a fully provisioned pool of servers at a quarter of its capacity by reading a scheduled event from a calendar and provisioning ahead of it. No reactive weight setting recovers this once the surge arrives. Free-text operator intents ground into control directives at $94\%$ exact match on previously unseen intents, and a single instruction reshapes live behavior end to end. We also quantify the latency, cost, and contention limits that bound it. The simulator, the proposed agentic framework, and the agent prompts are publicly available on \href{https://github.com/sotirischatzimiltis/AI-for-RAN-Resilience}{GitHub}. The supplementary material provides analytical \textit{proof} of the derived \textit{Theorem} on anticipation horizon, as well as the subsequent \textit{Corollary} results on reactive provisioning limits.
\end{abstract}


\section{Introduction}
Building on the foundations of open radio access networks (Open RAN)~\cite{oran_architecture}, AI-RAN advances the evolution toward AI-native operation by enabling higher levels of automation and orchestration, and the co-execution of radio access network (RAN) and AI workloads on shared edge infrastructure. As a result, the RAN is evolving from a communication-centric platform into an intelligent, compute-aware, and dynamically orchestrated system~\cite{ai-ran_nvidia, airan_alliance_vision}.

In this setting, resilience becomes a fundamental requirement that extends beyond traditional redundancy and fault tolerance~\cite{network_resiliences_a_systematic_approach}, since in Open RAN and AI-RAN  service continuity depends on coordinated behavior across disaggregated components,  open interfaces, and control loops operating at different timescales~\cite{resilience_and_failure_in_next_gen_comms_networks}. Network resilience is thus best understood not as a single mechanism but as a lifecycle. The literature decomposes it into four interdependent stages: \textit{anticipation} of emerging threats, \textit{absorption} of their impact while acceptable service is preserved, \textit{adaptation} under degraded or adversarial conditions, and \textit{recovery} of stable operation~\cite{farooq2025cyberresiliencenextgenerationnetworks, bennis2025resilientnativeintelligentnextgenerationwireless, resilient_by_design, reifert2024resiliencecriticalitybrothersarms}. We propose a fifth, \textit{evolution}, capturing what a system carries forward from one disruption to the next.

Existing literature has made significant progress on individual resilience functions across the anticipation, absorption, adaptation, and recovery stages. Predictive analytics and digital twins support anticipation~\cite{kasuluru_2025,masaracchia_2023}. Anomaly detection, fault isolation, and access control mechanisms support absorption~\cite{chatzimiltis_xai_llm_2026}. Resource allocation, virtual network function provisioning, and network slicing support adaptation~\cite{resource_allocation_network_slicing_recilience}, and self-healing mechanisms support recovery. However, these capabilities are typically studied as stage-specific or mechanism-specific solutions, with limited attention to how they can be systematically related and orchestrated across the disaggregated control hierarchy of Open RAN and AI-RAN. This paper addresses that gap by treating resilience as one continuous lifecycle and examining agentic orchestration as a unifying layer for coordinating its stages.

Agentic AI is well suited to this role. In contrast to isolated predictive or reactive models, it emphasizes goal-directed behavior, contextual reasoning, planning, tool use, and interaction among specialized agents. These properties are well aligned with resilience management, where decisions must integrate observations across multiple layers, reconcile local and global objectives, and sequence actions across heterogeneous network functions and timescales. In the context of AI-RAN, agentic AI is therefore better viewed as an orchestration layer over the intelligence hierarchy than as another analytics capability~\cite{elkael2026agentranagenticaiarchitecture, wu_2025, navidan_2026}. These capabilities, however, do not come for free. Embedding large language model (LLM)-based agents in closed-loop RAN control adds signaling, computation, and coordination overhead, and introduces new failure modes such as hallucinated or unsafe actions and adversarial manipulation of agent reasoning, all of which must be contained during deployment.

To ground this discussion, we consider signaling storm attacks in next-generation RAN as a motivating study. This scenario is particularly relevant because signaling storms can arise from compromised devices, bot-driven activity, or mass attachment scenarios, and can simultaneously disrupt control-plane stability, radio resource availability, and service continuity~\cite{signaling_storm_oran,samara_signaling_storm}. More importantly, the scenario naturally spans the full resilience lifecycle. As such, the study provides a concrete setting for illustrating how agentic orchestration can unify otherwise fragmented resilience mechanisms into a continuous control loop.

\noindent\textbf{Contributions.} 
\vspace{-0.1\baselineskip}
\begin{itemize}
\item \textbf{A multi-timescale agentic resilience orchestration algorithm.} We propose a three-tier algorithm in which an intent orchestrator at the service management and orchestration (SMO) layer, a tool-augmented non-real-time (Non-RT) agent, and a deterministic near-real-time (Near-RT) control loop coordinate through a shared policy. The design combines contextual perception, reasoning, cross-episode memory, and hierarchical coordination while deliberately keeping probabilistic LLM inference outside the fast actuation path.

\item \textbf{Analytical guarantees on the value of anticipation.} We derive and prove a sufficient minimum anticipation horizon under delayed capacity provisioning that guarantees the required capacity is online before a fast-onset surge and prevents backlog formation. We further prove that, when $\lambda_s > c_{\mathrm{ev}}\,\mu_{\mathrm{eff}}$ and $\tau_{\mathrm{prov}} > 0$, no purely reactive capacity policy can avoid a strictly positive backlog during the provisioning ramp, irrespective of its optimization objective or controller weights. This identifies operating conditions under which only anticipation, informed by external context, can prevent service loss.

\item \textbf{A signaling-storm resilience model and discrete-event AI-RAN simulator.} We formulate resilient control under benign and malicious signaling surges and develop a discrete-event simulator\footnote{The simulator, the proposed agentic framework, and the agent prompts are available at \url{https://github.com/sotirischatzimiltis/AI-for-RAN-Resilience}. The supplementary material contains the proofs of the anticipation-horizon theorem and the reactive-provisioning corollary.} that jointly captures retry-driven amplification, shared-compute contention, and non-instantaneous capacity provisioning. Performance is scored through service quality, capacity efficiency, and the anticipation, absorption, adaptation, and recovery over time (A3RT) resilience metric.

\item \textbf{An end-to-end evaluation of agentic reasoning and orchestration.} We evaluate several LLMs as the Non-RT agent, and compare the proposed approach with static, reactive Lyapunov, and rule-based controllers. We test reserve sizing on real scheduled events, show how provisioning delay and hidden compute contention limit every controller, and evaluate free-text intent grounding, compounded stressors, and multi-site orchestration. Throughout, we report resilience and benign attached rate alongside capacity efficiency, inference latency, and monetary cost, and we analyze the signaling overhead the agent adds.
\end{itemize}

\noindent\textbf{Organization.} The remainder of the paper develops the signaling-storm study and simulator (Section~\ref{sec:case-study}), presents the three-tier agentic framework (Section~\ref{sec:framework}), derives the analytical limits of provisioning (Section~\ref{ssec:limits}), evaluates the framework (Section~\ref{sec:analysis}), and concludes (Section~\ref{sec:conclusion}).

\section{Signaling-Storm System Model and Resilience Formulation}
\label{sec:case-study}

As a running study, we adopt a mass-attachment signaling storm in AI-RAN. It fits the purpose because it spans the full resilience lifecycle. Unusual signaling patterns enable anticipation, containing excess or malicious load supports absorption, adjusting processing capacity enables adaptation, draining the backlog constitutes recovery, and improving the response across repeated storms reflects evolution. 

\subsection{Scenario, Threat Model, and Service Rate}\label{ssec:sat}
We consider a mass-attachment scenario in which many user equipments (UEs) attempt to attach within a short interval, as in large-scale reconnection events or adversarial activity. We focus on UE initial attachment, since each attachment triggers a fixed exchange of Radio Resource Control (RRC) signaling and per-UE processing at the base station. The random-access preamble and response stay local to the radio unit (RU) and distributed unit (DU) below RRC, while the RRC Setup Request, Setup, and Setup Complete give $M=3$ control-plane messages that must reach the central unit (CU) where RRC terminates~\cite{3gppts38.331}.

We model attachment as an $M/M/c$ system of $c$ parallel servers at service rate $\mu$ with arrivals at rate $\lambda(t)$, stable only while $\lambda(t) < c\mu$. We define a signaling storm as an interval $[t_0,t_d]$, where $t_0$ and $t_d$ denote the storm onset and termination times, respectively, such that
\begin{equation}
    \lambda(t) > c\mu \quad \text{sustained over } [t_0, t_d],
    \label{eq:storm-condition}
\end{equation}
so requests accumulate faster than they get served.

\emph{Threat model.} The surge in attachment requests may be non-malicious, arising from mass reconnection after an outage or a large gathering, or adversarial. We accordingly split the arrival rate into a benign and a malicious component, $\lambda(t) = \lambda_{\text{ben}}(t) + \lambda_{\text{mal}}(t)$. In the adversarial case, an attacker drives compromised or bot-driven UEs that flood the control plane with legitimate-looking attach requests to deny service, without otherwise breaking the signaling protocol order. Benign surges must be absorbed and recovered from, whereas the malicious component can be identified and rate-limited at admission, motivating distinct responses. Flagging which arrivals are malicious is the task of an anomaly or intrusion detector, which we take as given and out of scope~\cite{chatzimiltis_xai_llm_2026}, so we study the response rather than the detection. A deployed detector is imperfect. We therefore assume that a fraction $\alpha_{\text{fp}}$ of legitimate attach attempts is wrongly flagged as malicious, so whenever the filter is active, these are dropped along with the botnet traffic.

\emph{Crowd-to-load mapping.} For the scheduled-event surges used in the evaluation, we relate a real crowd to a control-plane load. An event of attendance $N$ does not attach as one instantaneous burst. Only a fraction $\phi$ of attendees are active subscribers on the modeled cell, and their attempts spread over a short reconnection window $T_{\text{rw}}$ around the event boundary, giving
\begin{equation}
    \lambda_{\text{ben}}^{\text{rw}} = \frac{\phi\,N}{T_{\text{rw}}}.
    \label{eq:crowd-load}
\end{equation}
With $\phi \approx 0.2$ and $T_{\text{rw}} \approx 60$~s this calibrates to $\lambda_{\text{ben}}^{\text{rw}} \approx N/300$, so a crowd of $83{,}664$ maps to about $279$~UEs/s. This is roughly $60\%$ of the full-pool capacity $c_{\max}\mu \approx 459$~UEs/s ($c_{\max}=16$, with $\mu$ from~\eqref{eq:service-rate}), but far beyond what the few servers online at rest can serve, so it constitutes a storm. Because the same mapping converts both the true attendance and every controller's estimate into load, its constant does not favor any controller and only sets the load scale, which we fix so that the largest events approach $c_{\max}\mu$.

\emph{Service rate.} The service rate $\mu$ follows from the open interfaces and per-node processing. Each of the $M$ RRC messages crosses the disaggregated fronthaul and midhaul with a one-way delay $D_{\text{oneway}} = D_t + D_p + D_q + D_r$, the transmission, propagation, queuing, and processing delays along the O-RAN open fronthaul (O-FH) and the F1 control-plane (F1-C) midhaul. Combining this transport across all $M$ messages with the internal CU/DU processing $T_p = \sum_{i=1}^{M} t_{p,i}$ gives the mean attachment service time and per-server service rate
\begin{equation}
    \mu = \frac{1}{T_s}, \qquad T_s = T_p + M\,D_{\text{oneway}}.
    \label{eq:service-rate}
\end{equation}
The O-FH, based on the O-RAN 7.2x split, falls in the $\sim$100\,\textmu s class~\cite{oran_wg9_xtrp_req_v2}, and the split-option-2 F1-C carries a $1.5$--$10$~ms one-way
budget~\cite{3gpp_ts38801_v1400} whose lower bound, added to the O-FH delay, gives $D_{\text{oneway}}=1.6$~ms, the delay that disaggregation adds on top of $T_p$. 

\subsection{Discrete-Event Storm Dynamics}\label{ssec:dynamics}
The $M/M/c$ model of Section~\ref{ssec:sat} fixes arrivals externally and holds the service rate constant, so it misses three effects that dominate real storms and motivate a discrete-event model. The first is retry-driven amplification, where timed-out attempts re-enter as fresh load. The second is load-dependent service time, as the shared compute pool slows under contention. The third is non-instantaneous capacity provisioning, as new servers take time to come online. With all three disabled, the discrete-event model reduces to the analytical rate~\eqref{eq:service-rate}, so the two agree in the baseline case.

\emph{Retry-driven amplification.} Each benign attach attempt is governed by an RRC connection-setup guard timer $T_{300}$ (set at $1000$~ms). An attempt not served within $T_{300}$ is abandoned and retried after a random backoff (up to $500$~ms), up to $K$ attempts (set to 5 in the analysis), after which the UE fails to attach. A botnet UE is more aggressive, abandoning and re-attaching on a much shorter impatient period $T_b \ll T_{300}$, so its attempts accumulate far faster and dominate the amplification. Let $n(t) \in [1, K]$ be the average number of attempts a UE makes before it attaches or exhausts its retries. Since each UE generates $n(t)$ attempts rather than one, the effective load is
\begin{equation}
    \lambda_{\text{eff}}(t) = n(t)\,\lambda(t).
    \label{eq:lam-eff}
\end{equation}
Under light load $n(t) \approx 1$ and $\lambda_{\text{eff}} \approx \lambda$, whereas under overload, attempts wait past $T_{300}$ and retry, so $n(t) \approx K$ and $\lambda_{\text{eff}} \approx K\lambda$. The result is a self-reinforcing cycle in which longer queues cause more timeouts, and more retries lengthen the queues further.

\emph{Shared-compute contention.} In AI-RAN, the virtualized CU (vCU) and virtualized DU (vDU) processing runs on a finite compute pool shared with other workloads, so its speed depends on the pool load. Let $b$ be the number of busy attach-workers in a pool of size $\kappa = c_{\max}$, giving the compute occupancy $\omega = b/\kappa$, the fraction of the pool in use. A shared pool runs undisturbed while it has headroom and only contends once heavily loaded, so the processing time is the same up to a threshold $\omega_{\text{th}}$ and then rises linearly,
\begin{equation}
    s(\omega) =
    \begin{cases}
        1, & \omega \le \omega_{\text{th}},\\[4pt]
        1 + a\,\dfrac{\omega - \omega_{\text{th}}}{1 - \omega_{\text{th}}}, & \omega > \omega_{\text{th}},
    \end{cases}
    \label{eq:slowdown}
\end{equation}
with severity $a$, so the slowdown is bounded by $1+a$ rather than diverging. We assume that only the processing part of the service time is affected, while the transport delay $M\,D_{\text{oneway}}$ stays fixed,
\begin{equation}
    T_s^{\text{eff}} = s(\omega)\,T_p + M\,D_{\text{oneway}},
    \qquad
    \mu_{\text{eff}} = \frac{1}{T_s^{\text{eff}}}.
    \label{eq:contention}
\end{equation}
Below the threshold $s = 1$ and $\mu_{\text{eff}} = \mu$, recovering the uncontended rate, and above it each busy worker slows every attach a little more. Because the pool size is $\kappa = c_{\max}$, contention begins only when most servers are busy ($\omega_{\text{th}}=0.65$), which happens when the controller provisions heavily. A controller that plans with the nominal $\mu$ therefore assumes each server is faster than it is and under-provisions precisely when it is relying on that capacity most.

\emph{Capacity provisioning.} A controller commands a target capacity $c^{\star}(t)$, but the online capacity $c(t)$ tracks it with delay. Bringing additional attachment-processing capacity into service takes $\tau_{\text{prov}}$ seconds and is assumed to  be added serially. Scaling down is immediate, as a server simply stops taking new work without interrupting any in-flight attach. Thus
\begin{equation}
    c(t) =
    \begin{cases}
        c^{\star}(t), & c^{\star}(t) \le c(t-1),\\[4pt]
        c(t-1) + 1 \ \text{per } \tau_{\text{prov}}, & c^{\star}(t) > c(t-1).
    \end{cases}
    \label{eq:provision}
\end{equation}
This delay is what gives anticipation its value. With $\tau_{\text{prov}} > 0$, capacity added only after a surge begins arrives too late to prevent the backlog, so provisioning ahead of the surge materially changes the outcome.

\emph{Traffic and queue evolution.} Arrivals follow a specific profile, $\lambda(t)$  rises from a calm baseline during a storm, as a step or a ramp, and then returns to it. With $L_q(t)$ the number of attempts waiting for a server, the queue evolves as
\begin{equation}
    L_q(t{+}1) = \max\!\Big(0,\;
    L_q(t) + \lambda_{\text{eff}}(t) - c(t)\,\mu_{\text{eff}}(t)\Big),
    \label{eq:queue}
\end{equation}
which brings together the retry-amplified load $\lambda_{\text{eff}}(t)$, the contention-degraded rate $\mu_{\text{eff}}(t)$, and the delayed capacity $c(t)$. Equation~\eqref{eq:queue} is the fluid average, which the simulator realizes stochastically by modeling each UE and its retries.

\subsection{Resilience Quantification}\label{ssec:resilience_quant}
To score system health over a storm we define a utility function $u(t) \in [0,1]$, reported as a percentage, where higher values indicate a more stable, better-served system. It combines a capacity-utilization term and a queue-length term, each a logistic (sigmoid) function that stays near $100\%$ under light conditions and falls toward zero as the system is stressed,
\begin{align}
    u(t) &= w_A\,u_A(t) + w_B\,u_B(t), \qquad w_A + w_B = 1,
    \label{eq:utility}\\
    u_A(t) &= \frac{1}{1 + e^{\,k_A(\rho(t) - m_{\text{frac}A})}},
    \quad \rho(t) = \frac{\lambda(t)}{c(t)\,\mu},
    \nonumber\\
    u_B(t) &= \frac{1}{1 + e^{\,k_B(L_q(t) - m_B)}},
    \quad m_B = L_{q,\max}\,m_{\text{frac}B},
    \nonumber
\end{align}
where $\rho(t)$ is the utilization, the ratio of offered load $\lambda(t)$ to online serving capacity $c(t)\,\mu$, so $\rho < 1$ indicates spare capacity and $\rho \ge 1$ saturation. The constants $k_A, k_B$ set the steepness of each transition and $m_{\text{frac}A}, m_B$ their knees. $u_A$ passes through $0.5$ at the utilization knee $\rho = m_{\text{frac}A}$, retaining near-full utility while the cell runs comfortably below capacity and falling toward zero as it approaches and exceeds saturation, while $u_B$ falls as the queue nears its overflow threshold $L_{q,\max}$. Defining $u_A$ through utilization rather than the raw arrival rate keeps it independent of pool size. The same parameters $(k_A, m_{\text{frac}A})$ therefore apply whether few or many servers are online, whereas a raw-rate form would need retuning as $c(t)$ changes. Utility thus measures load relative to current capacity, staying near $100\%$ under normal load and degrading as the cell nears saturation or the queue builds.

We quantify resilience $P \in [0,1]$, reported as a percentage, over a disruption with the A3RT metric~\cite{reifert2024resiliencecriticalitybrothersarms}, which weights the utility retained during absorption and adaptation against a desired profile $u_{\text{des}}$ and adds a recovery-time term,
\begin{equation}
    P = w_1\frac{\int_{t_0}^{t_d} u\,dt}{\int_{t_0}^{t_d} u_{\text{des}}\,dt}
      + w_2\frac{\int_{t_d}^{t_r} u\,dt}{\int_{t_d}^{t_r} u_{\text{des}}\,dt}
      + w_3\,t_{\text{rec}},
    \label{eq:a3rt}
\end{equation}
\begin{equation}
    t_{\text{rec}} =
    \begin{cases}
        1, & t_r - t_0 \le \Delta t_{\text{des}},\\[2pt]
        \dfrac{\Delta t_{\text{des}}}{t_r - t_0}, & \text{otherwise},
    \end{cases}
    \qquad \textstyle\sum_i w_i = 1,
    \label{eq:trec}
\end{equation}
where $[t_0, t_d]$ is the storm interval and $t_r$ the recovery time, the first instant after $t_d$ at which utility returns above a fraction $\eta_{\text{rec}}$ of $u_{\text{des}}$ and holds there for at least $\Delta t_{\text{hold}}$ ($\eta_{\text{rec}} = 0.95$, $\Delta t_{\text{hold}} = 30$~s). We set $u_{\text{des}}$ to the mean pre-storm utility measured while the cell operates at its steady resting capacity, excluding any pre-provisioning ramp, so that $P$ measures resilience relative to the system's own healthy baseline rather than a transiently over-provisioned state.

\subsection{Controllers}\label{ssec:controllers}
The dynamics of Section~\ref{ssec:dynamics} expose two actuation points, the commanded capacity $c^{\star}(t)$ and the malicious-drop probability $p_{\text{drop}}$. We introduce three baseline controllers of increasing capability as reference points for the framework of Section~\ref{sec:framework}. The first two act on capacity alone, the third also actuates the admission filter through fixed rules.

\emph{Static.} The server count is held constant, $c^{\star}(t) = c$ for a chosen $c \in \{1,\dots,c_{\max}\}$. Using no feedback, it shows how a statically provisioned system copes with a storm.

\emph{Reactive.} A feedback controller resizes the pool at every control step, balancing queue growth against the cost of idle servers through a Lyapunov drift-plus-penalty objective $J(c)$. With $a_n = \lambda(t)/\mu - c$ the unmet demand in servers and $q_n = L_q(t)/L_{q,\max}$ the normalized backlog, it selects
\begin{equation}
\begin{split}
    c^{\star}(t) &= \arg\min_{1 \le c \le c_{\max}} J(c), \\
    J(c) &= q_n\,a_n + \tfrac{1}{2}\,a_n^2 - V\,u_c(t) + W\,\frac{c}{c_{\max}},
\end{split}
\label{eq:lyapunov}
\end{equation}
where $u_c(t)$ is the utility of~\eqref{eq:utility} evaluated at $c$. The drift terms $q_n\,a_n$ and $\tfrac{1}{2}\,a_n^2$ push toward more servers when the backlog is long or arrivals outpace capacity, $-V\,u_c(t)$ rewards high utility, and $W\,c/c_{\max}$ discourages over-provisioning. Both terms lie in $[0,1]$, so $V$ and $W$ are directly comparable and default to $1$.

\emph{Rule-based.} Our strongest baseline replaces the agent's language-model policy with a fixed decision tree over the same telemetry, isolating the value of language-model reasoning. Like the agentic framework, it runs on two timescales. A slow rule periodically sets the policy, and the reactive update~\eqref{eq:lyapunov} applies it at every control step. The rule flags the cell as elevated when $\lambda(t)$ rises clearly above its resting baseline and, lacking a calendar, treats every elevated load as malicious. It then raises $V$ for headroom and engages the filter with a drop probability that grows with overload severity, hardening if utility keeps falling. It can also raise $V$ early when the forecast extrapolates an already rising trend. This is close to the best that hand-tuned rules can achieve on this telemetry, but its limit is structural, since it responds only once arrivals begin to climb, never before.

Admission filtering is therefore imperfect. Dropping a fraction $p_{\text{drop}}$ of malicious attempts also removes benign traffic at rate $\alpha_{\text{fp}}\,p_{\text{drop}}$, so containment trades off against denying legitimate users, a cost the rule pays on every benign surge it cannot recognize. The framework of Section~\ref{sec:framework} addresses exactly these gaps, separating a scheduled surge from an attack and provisioning before the queue grows.

\subsection{Metrics}\label{ssec:metrics}
\emph{Per-window resilience.} A run may contain several storm windows of different character, so we evaluate $P$ of~\eqref{eq:a3rt} over each window $[t_0^{(k)}, t_r^{(k)}]$ separately and report the episode resilience as the mean of the per-window scores, $P = \frac{1}{N_w}\sum_{k=1}^{N_w} P^{(k)}$, so a controller cannot hide a starved window behind a healthy one. The evaluation scenario pairs a malicious botnet ramp with a benign event surge, giving two scores $P_{\text{bot}}$ and $P_{\text{surge}}$ that reward opposite responses, filtering the attack and provisioning for the crowd.

\emph{Service quality.} Resilience rewards utility held but not how many users were served. Each legitimate UE ends up attached, failed after exhausting its retries, or dropped at admission, with counts $A_b$, $F_b$, and $D_b$, giving the benign attached and false-positive rates
\begin{equation}
    S_b = \frac{A_b}{A_b + F_b + D_b}, \qquad
    \phi_b = \frac{D_b}{A_b + F_b + D_b}.
    \label{eq:benign-served}
\end{equation}
When the surge is entirely benign, $\phi_b$ measures pure over-filtering. For the attack we report the filtered fraction, the share of botnet UEs deliberately dropped out of all botnet outcomes $A_m$, $F_m$, and $D_m$,
\begin{equation}
    \Phi_m = \frac{D_m}{A_m + F_m + D_m},
    \label{eq:filtered}
\end{equation}
which counts only admission drops, so a controller that never filters stays at zero however overloaded.

\emph{Capacity cost and efficiency.} Resilience can always be bought with capacity, so we track its cost. With $\bar{c}$ the mean number of servers online, the capacity efficiency
\begin{equation}
    E = \frac{P}{\bar{c}/c_{\max}},
    \label{eq:efficiency}
\end{equation}
is the resilience per unit of provisioned capacity, normalized so $E=1$ marks a controller that holds all $c_{\max}$ servers to reach $P=1$. Since it rewards economy, an under-provisioned low-$P$ controller can also score high, so we read $E$ only alongside $P$.

\emph{Agentic overhead and grounding.} For the agentic controllers, we also record the per-decision latency and monetary cost of the language model, and we compare the agent's estimated event attendance against ground truth to confirm that its provisioning is grounded rather than fortunate.

\section{The Proposed Agentic Resilience Framework}\label{sec:framework}
\begin{figure}[!t]
    \centering \includegraphics[width=0.99\columnwidth]{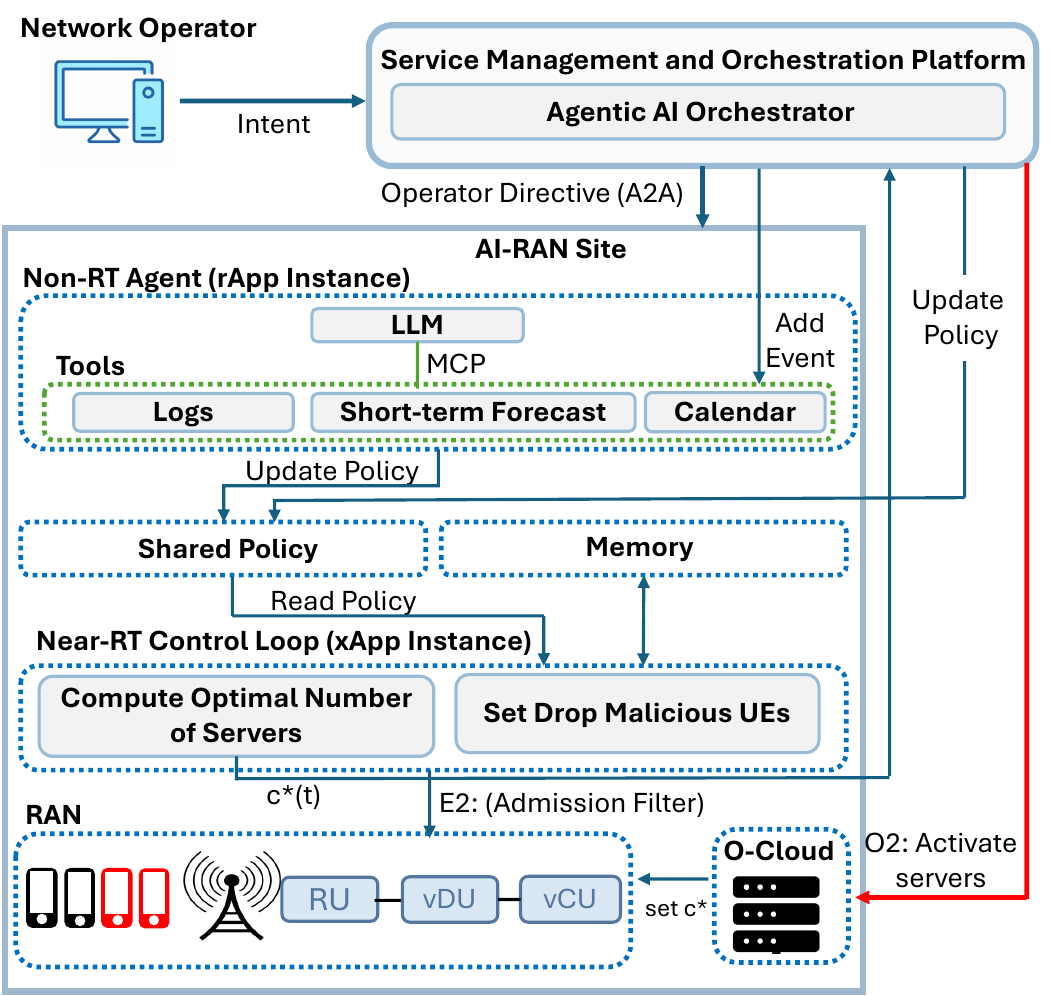}
    \caption{The Proposed Agentic Resilience Framework Architecture for the AI-RAN.}
    \label{fig:framework}
\end{figure}

Realizing resilience as a continuous lifecycle rather than a set of stage-specific mechanisms requires a framework that binds those stages into one coordinated cycle and carries what is learned from each disruption into the next. Agentic AI is a natural candidate, but placing LLM-based agents inside the real-time or Near-RT control loops adds signaling and decision latency exactly when the network is overloaded. We therefore propose an agentic resilience framework that coordinates the lifecycle across coupled loops running at different timescales while keeping slow, probabilistic LLM reasoning off the fast control path. Following the O-RAN control hierarchy, the framework spans three tiers, shown in Fig.~\ref{fig:framework}. An SMO-level orchestrator turns operator intent into standing policy, a per-site Non-RT agent judges storm conditions, and a deterministic Near-RT loop enforces those decisions at every control step.

\begin{algorithm}[!t]
\footnotesize
\caption{Near-RT Control Loop.}
\label{alg:near-rt-control}
\begin{algorithmic}[1]
\Require $\lambda(t)$, $\lambda_{\mathrm{rest}}$, $L_q(t)$, $c(t)$, $\pi$,
  $\tau_{\mathrm{prov}}$, $\beta_{\mathrm{rel}}$, optional cross-episode memory $\mathcal{B}$
\Ensure commanded capacity $c^{\star}(t)$, malicious-drop probability $p_{\mathrm{drop}}(t)$
\While{episode active} \Comment{every $T_{\mathrm{ctl}}$}
\State $\tilde{\pi} \gets$ atomic snapshot of $\pi$

  \Statex \quad $\triangleright$ \textit{Reserve timing: online BY the event, never earlier}
  \State $T_{\mathrm{ramp}} \gets (\tilde{\pi}.c_{\mathrm{res}} - 1)\,\tau_{\mathrm{prov}}$
    \Comment{$T^{\min}_{\mathrm{ant}}$ of~\eqref{eq:tant} at $c_{\mathrm{ev}}{=}1$}
  \State $\mathit{due} \gets \big(\tilde{\pi}.t_{\mathrm{ev}} \le 0\big)\ \textbf{or}\
    \big(t \ge \tilde{\pi}.t_{\mathrm{ev}} - T_{\mathrm{ramp}} - T_{\mathrm{ctl}}\big)$
  \State $\mathit{over} \gets \big(\tilde{\pi}.t_{\mathrm{ev}} > 0\big)\ \textbf{and}\
    \big(t > \tilde{\pi}.t_{\mathrm{ev}}\big)\ \textbf{and}\
    \big(\lambda(t) \le \beta_{\mathrm{rel}}\,\lambda_{\mathrm{rest}}\big)$
  \State $c_{\mathrm{res}}^{\mathrm{eff}} \gets \tilde{\pi}.c_{\mathrm{res}}$
    \textbf{if} ($\mathit{due}$ \textbf{and not} $\mathit{over}$) \textbf{else} $1$
  \State $\underline{c} \gets \max\{\tilde{\pi}.c_{\mathrm{floor}},\ c_{\mathrm{res}}^{\mathrm{eff}}\}$
    \Comment{operator floor $\cup$ scheduled reserve}

  \Statex \quad $\triangleright$ \textit{Capacity command under the current posture}
  \State $c^{\star} \gets \arg\min_{1 \le c \le c_{\max}} J(c)$ under $\tilde{\pi}.V,\ \tilde{\pi}.W$
  \State $c^{\star} \gets \min\{\max\{c^{\star},\ \underline{c}\},\ c_{\max}\}$
  \If{$L_q(t) > \tilde{\pi}.L_{\mathrm{hold}}$}
    \State $c^{\star} \gets \max\{c^{\star},\ c(t)\}$
      \Comment{no scale-down while the queue drains}
  \EndIf

  \Statex \quad $\triangleright$ \textit{Admission filter, gated by the storm verdict}
  \If{$\tilde{\pi}.\mathrm{storm}$}
    \State $p_{\mathrm{drop}} \gets \tilde{\pi}.p_{\mathrm{drop}}$
  \ElsIf{$\mathcal{B}$ enabled \textbf{and} the live signature matches an entry}
    \State $p_{\mathrm{drop}} \gets \mathcal{B}.p_{\mathrm{drop}}$
      \Comment{engage from memory before the agent rules}
  \Else
    \State $p_{\mathrm{drop}} \gets 0$
  \EndIf

  \State actuate $c^{\star}$; $c(t)$ tracks it with delay $\tau_{\mathrm{prov}}$
    under \eqref{eq:provision}
  \State drop each \emph{flagged} arrival with probability $p_{\mathrm{drop}}$
\EndWhile
\State \textbf{on} episode end \textbf{do} store the storm signature, its
  engagement level, the containing drop strength, and the tuned weights in  $\mathcal{B}$
\end{algorithmic}
\end{algorithm}

\subsection{Near-RT Control Loop}
The Near-RT loop is the framework's deterministic actuator, detailed in Algorithm~\ref{alg:near-rt-control}. It runs at a fixed control cadence and, at every step, reads the current attach arrival rate $\lambda(t)$, the queue length, and the number of servers online, together with an atomic snapshot of the current policy, which holds the storm verdict, the malicious drop probability, and the controller weights the agent may have set. From these, it solves the drift-plus-penalty objective~\eqref{eq:lyapunov} for the commanded capacity $c^\star(t)$. This command passes through two guardrails. It is clamped to the operator capacity floor and the physical maximum, and the loop refuses to scale down while the queue is still draining, which prevents it from oscillating during recovery. In parallel it applies the malicious drop probability $p_\mathrm{drop}$, gated by the agent's storm verdict, so admission filtering is active only when a storm is declared. Crucially, no language model sits on this path. The loop actuates on the last policy it received and never waits for the agent to respond, keeping decision latency off the fast control path even when the agent is slow or the network is overloaded. Each step is also computationally light, since every evaluation of $J(c)$ in~\eqref{eq:lyapunov} is $\mathcal{O}(1)$, the capacity search costs $\mathcal{O}(c_{\max})$, and the optional signature lookup in the cross-episode memory $\mathcal{B}$ costs $\mathcal{O}(|\mathcal{B}|\,d_{\mathrm{sig}})$ for $d_{\mathrm{sig}}$-dimensional signatures, giving $\mathcal{O}(c_{\max} + |\mathcal{B}|\,d_{\mathrm{sig}})$ time per control period and $\mathcal{O}(|\mathcal{B}|\,d_{\mathrm{sig}})$ memory. With $c_{\max}$ and $|\mathcal{B}|$ small in practice, the loop comfortably fits within its control period $T_{\mathrm{ctl}}$, which lies between $10$~ms and $1$~s. As shown in Fig.~\ref{fig:framework}, an xApp applies admission filtering over E2, while the Near-RT loop, which alone observes the per-second queue and arrival telemetry, passes $c^\star(t)$ to the SMO, which activates the servers on the O-RAN cloud (O-Cloud) over O2~\cite{ORAN_WG1_ARCH_DESC}. The resulting instantiation delay is captured by $\tau_{\mathrm{prov}}$, which we sweep in Section~\ref{ssec:vw_sweep}.

\subsection{Non-RT Agent}\label{ssec:nonrtagent}
The Non-RT agent is the per-site decision maker, summarized in Algorithm~\ref{alg:non-rt-agent}. Its input is the telemetry that the Near-RT loop of Algorithm~\ref{alg:near-rt-control} produces at every control step, and its output is a shared policy object that the same loop reads back at its next step. The two algorithms therefore form a closed loop, with Algorithm~\ref{alg:non-rt-agent} deciding and Algorithm~\ref{alg:near-rt-control} enforcing.  It runs on its own schedule every few seconds, apart from the fast control path, and it never acts on the network directly. At each assessment, it receives a summary of the recent telemetry together with its own previous verdict, shown in  Fig.~\ref{fig:nonrt_input}. The summary reports how the arrival rate, the queue, and the retry rate have moved over the last window. Each signal is also tagged as leading or lagging. Arrivals and retries rise at storm onset, whereas the queue reacts only afterwards and is held down by the fast loop, so the agent should base its judgment on the leading signals. Feeding the previous verdict back keeps the decision steady and stops the agent from switching the storm state on and off on brief noise. To look beyond raw arrivals, the agent can also call a small set of tools over the Model Context Protocol (MCP)~\cite{mcp_spec} that return the live episode logs, a short forecast of the arrival rate, and a calendar of planned events.

From this input, the agent returns one structured decision, shown in Fig.~\ref{fig:trace}. In this example, arrivals are calm and no storm is present, but a call to the calendar reveals an upcoming event, so the agent estimates the crowd and the time of its surge, and the system turns these into a reserve provisioned just in time. When a storm is present instead, the same schema carries the storm verdict, a drop probability scaled to the overload severity, and controller weights that add capacity headroom.

\begin{algorithm}[!t]
\footnotesize
\caption{Non-RT Agent (site-local, period $T_{\mathrm{asm}}$, asynchronous).}
\label{alg:non-rt-agent}
\begin{algorithmic}[1]
\Require telemetry window $\mathcal{H}$, previous policy $\pi^{-}$, $\pi.\mathrm{instr}$
\Ensure candidate policy $\pi^{+}$, published into $\pi$
\While{\textsc{true}} \Comment{every $T_{\mathrm{asm}}$}
\State extract arrival, retry, and queue trajectories from $\mathcal{H}$
  \State recall the verdict in $\pi^{-}$ to keep storm state steady
  \State $\sigma \gets \Call{LogTool}{\,}$,\ $\hat{\lambda} \gets \Call{ForecastTool}{\,}$,\
    $\mathcal{E} \gets \Call{CalendarTool}{\,}$
  \Statex \quad $\triangleright$ \textit{Anticipation: commit a reserve once for an upcoming event}
  \If{$\mathcal{E}$ names an upcoming event not yet provisioned}
    \State $\hat{N} \gets$ attendance inferred from the venue and context in $\mathcal{E}$
    \State $\pi^{+}\!.c_{\mathrm{res}} \gets$ reserve holding $\hat{N}$ below the target utilization
    \State $\pi^{+}\!.t_{\mathrm{ev}} \gets \mathcal{E}.t$ \Comment{fast loop brings it online just in time}
  \EndIf
  \Statex \quad $\triangleright$ \textit{Classification: judge the current load, honoring any delegation}
  \If{the latest arrival rate sits clearly above rest}
    \If{an event in $\mathcal{E}$ is due now ($t \ge \mathcal{E}.t$), \textbf{or}
        $\pi.\mathrm{instr}$ marks this load benign}
      \State $\pi^{+}\!.\mathrm{storm} \gets \textsc{false}$,\
        $\pi^{+}\!.p_{\mathrm{drop}} \gets 0$ 
    \Else
      \State $\pi^{+}\!.\mathrm{storm} \gets \textsc{true}$,\
        $\pi^{+}\!.p_{\mathrm{drop}} \gets$ scaled to overload severity 
      \State $\pi^{+}\!.V \gets$ raised for capacity headroom
    \EndIf
  \Else
    \State $\pi^{+}\!.\mathrm{storm} \gets \textsc{false}$; return $V$ toward default once load is at rest
  \EndIf
  \State write $\pi^{+}$ into $\pi$ atomically, operator overrides left in force
\EndWhile
\end{algorithmic}
\end{algorithm}

\begin{figure}[!t]
\begin{tcolorbox}[colback=white, colframe=black, sharp corners,
                  boxrule=0.5pt, width=\columnwidth]
\small
\textbf{Non-RT Agent Input} \hfill \textit{Assessment \#12, $t = 200$\,s}
\begin{itemize}
    \item \textbf{Current policy} (updated 5\,s ago): \\
    \texttt{storm\_active} = false \\
    \texttt{malicious\_drop\_prob} = 0.00 \\
    \texttt{reserve\_servers} = 1 \\
    \texttt{queue\_hold\_threshold} = 10 \\
    \texttt{lyapunov\_V} = \texttt{lyapunov\_W} = 1.0
    \item \textbf{Telemetry} (last 15\,s, 16 samples): \\
    Arrival rate $\lambda$: 21 $\rightarrow$ 22 UEs/s (flat) \\
    peak $\lambda$: 25 at $t{=}193$\,s \textit{(leading)} \\
    $\lambda$ baseline: $\approx$20 UEs/s \\
    $\lambda$ bin means: 21, 20, 22, 21, 20, 22, 21, 22 \\
    Retry rate: 3/s $\rightarrow$ 3/s (flat) \textit{(leading)} \\
    Queue length: 0 $\rightarrow$ 0 (flat) \textit{(lagging)}
\end{itemize}

\textbf{Task}
\begin{itemize}
    \item \emph{``Judge storm-vs-noise from the window trends and return a PolicyUpdate.''}
\end{itemize}
\end{tcolorbox}

\caption{Non-RT agent input at a single assessment.}\label{fig:nonrt_input}
\end{figure}

\begin{figure}[!t]
\begin{tcolorbox}[colback=white, colframe=black, sharp corners,
                  boxrule=0.5pt, width=\columnwidth]
\small
\textbf{Non-RT Agent Output}
\begin{itemize}
    \item \textbf{Storm state:} \\
    \texttt{storm\_active} = false \\
    \texttt{malicious\_drop\_prob} = 0.00
    \item \textbf{Control parameters:}\\
    \texttt{tighten} = true \\
    \texttt{lyapunov\_V} = \texttt{lyapunov\_W} = 1.0 \\
    \texttt{queue\_hold\_threshold} = 10
    \item \textbf{Expected load:} 82{,}000 attendees at $t = 240$\,s.
    \item \textbf{Rationale:} \emph{``England v Brazil at Wembley, a 90k-seat venue, not sold out; a marquee fixture drawing a large but sub-capacity crowd.''}
\end{itemize}

\textbf{Reasoning}
\begin{itemize}
    \item \emph{``Load is at rest with no storm. A scheduled surge is due at $t = 240$\,s, so pre-provision a reserve for the estimated crowd.''}
\end{itemize}
\end{tcolorbox}

\caption{Non-RT agent output for pre-event provisioning.}\label{fig:trace}
\end{figure}

Because the agent only decides while the deterministic loop enforces, its slow and uncertain reasoning shapes the control posture without ever reaching the Near-RT actuation path.

\subsection{Agent Tools}
Left to telemetry alone, the agent sees only arrivals and cannot tell a malicious flood from a legitimate surge, nor act before load arrives. Three read-only tools close these gaps.

\begin{itemize}
\item \textbf{Log tool ($\sigma$).} It reports the live episode, returning cumulative service and security counters together with a running measure of how well utility is being held against its calm pre-storm baseline. This measure is computed only from the telemetry seen so far, with no knowledge of the future, and it tells the agent whether its current filtering is coping or needs to push harder.
\item \textbf{Forecast tool ($\hat\lambda$).} It returns a short prediction of where load is heading over the next few control steps, projecting the arrival rate along with the retry rate and the queue. The prediction is a least-squares trend fitted to the recent telemetry window, with the fit quality reported as a confidence score so the agent can discount a poor fit. When the trend is rising, the agent can raise the headroom weight $V$ a few cycles early, so the Near-RT loop begins adding servers before the queue builds. The forecast thus complements the calendar by flagging unscheduled ramps as they emerge, while reserves for scheduled crowds remain the calendar's responsibility.

\item \textbf{Calendar tool ($\mathcal{E}$).} It exposes scheduled load events that the SMO orchestrator has registered or a calendar sub-agent has scraped from the web, such as a stadium egress or a planned mass registration. Each event is named with its venue and whether it sold out, together with the time its surge is expected, but never the turnout. The agent estimates the crowd itself from the event and venue, and the system turns that estimate into a reserve provisioned ahead of the surge. Because the surge is on the calendar, the agent classifies it as benign and admits it without filtering, so no legitimate user is dropped during the event.
\end{itemize}

\subsection{Shared Policy}
The two loops never call each other. They communicate through a single shared policy object that the agent writes and the Near-RT loop reads, which lets them run at different speeds. The agent updates the policy whenever it finishes an assessment, while the fast loop reads a complete copy of it at every control step and never sees a half-written update. The policy holds four kinds of fields. The first are the operational levers, the storm verdict and the malicious drop probability, which are refreshed on every assessment. The second are the slower controller settings, the utility and cost weights and the queue-hold threshold, which change only when the agent decides to retune, so the controller does not keep shifting on every cycle. The third is the scheduled reserve, a number of servers to hold online ahead of a planned surge together with the time the surge is due, which the agent commits once for an event and the fast loop brings online just before it starts. The fourth are operator overrides set by the SMO orchestrator, such as a fixed weight posture or a minimum capacity floor, which take priority over the agent's own tuning. Because the loop reads only this object, it keeps working correctly no matter when, or even whether, the agent updates it. If the agent fails, the loop can fall back to the rule-based controller of Section~\ref{ssec:controllers}, which drives the same actuators.

\subsection{Agent Memory}
The mechanisms above let the framework react and anticipate within a single storm. A further layer lets it improve across storms. The fast loop keeps a lightweight memory of each storm's signature, its baseline rate, the level at which it engages, and the drop strength that contained it. Once a pattern is recognized, it can turn on the admission filter before the agent has ruled. This memory, along with the tuned controller settings, is carried across episodes, so a site that has already weathered a storm starts the next one configured for it rather than from defaults. Only this slow-moving state is kept. The live levers, which are decisions about the present, are always recomputed. Memory is disabled in all experiments of Section~\ref{sec:analysis} to isolate within-storm behavior, and its evaluation is left to future work.

\begin{algorithm}[!t]
\footnotesize
\caption{SMO Orchestrator (event-driven).}
\label{alg:agentic_orchestrator}
\begin{algorithmic}[1]
\Require operator intent $I$ (free text), site policy $\pi$
\Ensure updated overrides in $\pi$, calendar entry for a scheduled event, standing instruction $\pi.\mathrm{instr}$
\Procedure{OnIntent}{$I$} \Comment{triggered on intent arrival}
    \State $d \gets \Call{Ground}{I}$ \Comment{map intent to directive over the five levers}
    \If{$d.\mathrm{posture} \neq \varnothing$}
        \State $(\pi.V, \pi.W) \gets \Call{PostureWeights}{d.\mathrm{posture}}$
    \EndIf
    \If{$d.\mathrm{floor} \neq \varnothing$}
        \State $\pi.c_{\mathrm{floor}} \gets d.\mathrm{floor}$
    \EndIf
    \If{$d.\mathrm{event} \neq \varnothing \land d.\mathrm{event}.t \neq \varnothing$}
        \State \Call{Calendar.Register}{$d.\mathrm{event}$}
    \EndIf
    \If{$d.\mathrm{instr} \neq \varnothing$}
        \State $\pi.\mathrm{instr} \gets d.\mathrm{instr}$ \Comment{delegate to Non-RT agent}
    \EndIf
\EndProcedure
\end{algorithmic}
\end{algorithm}

\begin{table*}[t]
\centering
\caption{Non-RT agent LLM comparison over five seeds (mean $\pm$ 95\% CI).}
\label{tab:exp1_llm}
\setlength{\tabcolsep}{3.5pt}
\scriptsize
\resizebox{\textwidth}{!}{%
\begin{tabular}{lcccccccccc}
\toprule
\multirow{2}{*}{Model} & Overall Resil. (\%) & Bot Resil. (\%) & Surge Resil. (\%) & Benign Attached (\%) & Crowd Est. & Reserve Servers & Bot Filtered (\%) & Mean Capacity & Latency/assess. & Cost/assess. \\
 & $P\uparrow$ & $P_{\text{bot}}\uparrow$ & $P_{\text{surge}}\uparrow$ & $S_b\uparrow$ & ($\times 10^3$, true 83.7) & (srv, ideal 13) & $\Phi_m\uparrow$ & $\bar{c}$ (srv) & (s)\,$\downarrow$ & (\$)\,$\downarrow$ \\
\midrule
\textbf{GPT-5.4-Mini (on)} & $89.9\pm1.1$ & $83$ & $97$ & $99.8\pm0.1$ & $82.2\pm3.3$ & $12.4\pm0.7$ & $33\pm29$ & $4.1\pm0.2$ & $23.2\pm3.0$ & $0.0123$ \\
QWEN3.7-Plus               & $87.5\pm0.6$ & $84$ & $92$ & $99.7\pm0.0$ & $68.0\pm8.3$ & $10.2\pm1.0$ & $73\pm2$ & $3.7\pm0.1$ & $5.5\pm0.2$ & $0.0026$ \\
\textbf{GEMINI-3.1-Flash-Lite} & $87.4\pm1.0$ & $84$ & $91$ & $99.7\pm0.1$ & $69.0\pm10.2$ & $10.2\pm1.4$ & $84\pm3$ & $3.6\pm0.2$ & $3.2\pm1.1$ & $0.0018$ \\
GPT-5.4-Mini (off)         & $87.1\pm2.0$ & $84$ & $91$ & $98.9\pm1.1$ & $81.4\pm12.7$ & $12.4\pm2.1$ & $84\pm4$ & $3.6\pm0.3$ & $6.3\pm0.7$ & $0.0060$ \\
\midrule
CLAUDE-HAIKU-4.5           & $76.7\pm6.7$ & $84$ & $70$ & $50.4\pm34.2$ & $48.0\pm8.3$ & $7.4\pm1.1$ & $77\pm5$ & $3.5\pm0.0$ & $5.2\pm0.1$ & $0.0103$ \\
GPT-4o-Mini                & $71.6\pm2.6$ & $84$ & $60$ & $35.4\pm3.1$ & $14.0\pm27.2$ & $2.2\pm4.3$ & $71\pm5$ & $3.0\pm0.5$ & $4.5\pm0.7$ & $0.0010$ \\
\bottomrule
\multicolumn{11}{@{}l}{\scriptsize\textit{Note:} (on)/(off) = reasoning enabled/disabled.}\\
\end{tabular}%
}
\end{table*}

\subsection{Agentic AI Orchestrator}\label{ssec:orchestrator}
The orchestrator is the topmost tier and the interface between the human operator and the sites, shown in Algorithm~\ref{alg:agentic_orchestrator}. Its outputs feed both lower tiers. The posture and floor it writes are read by the Near-RT loop of Algorithm~\ref{alg:near-rt-control}, while the calendar entry and standing instruction are consumed by the Non-RT agent of Algorithm~\ref{alg:non-rt-agent}. Sitting at the SMO above the per-site agents, and it is event-driven, staying idle while the network runs normally and activating only when an operator issues an intent, so it adds no steady-state signaling or real-time latency. Its role is to turn free-text operator intents into a single standing directive that the lower tiers apply without further review. The directive carries five levers, the posture direction and an optional graded strength that together set the utility and cost weights, a hard capacity floor, a scheduled load event with its venue, sold-out status, and time, and a standing instruction that tells the site agent how to interpret conditions rather than change capacity. The orchestrator writes the posture and the floor into the shared policy as operator overrides, which take priority over the site agent's own tuning, registers the named event on the calendar the site agent reads, and leaves the standing instruction for it to apply on every assessment. 

Sitting at the SMO above the sites, the orchestrator coordinates one or many of them from a single point. It decomposes one operator intent into heterogeneous per-site directives and dispatches them over the Agent2Agent (A2A) protocol~\cite{a2a_github}, so a scheduled stadium egress raises the reserve only on the cells that serve it while a fleet-wide cost posture applies everywhere, reconciling one global objective with the local conditions of each site. Each site then runs its own Non-RT agent and Near-RT loop against the directive it receives, so coordination stays centralized at the SMO while enforcement stays local and independent.

\section{Analytical Limits of Provisioning}\label{ssec:limits}
The dynamics of Section~\ref{ssec:dynamics} and the controllers of Section~\ref{ssec:controllers} together determine the lead time required for capacity provisioning to absorb a fast-onset surge without backlog. The result below gives this minimum anticipation horizon under the fluid-queue model of~\eqref{eq:queue}, which the discrete-event simulator realizes stochastically. Here, $c(t)$ denotes the number of servers online at time $t$, $\mu_{\text{eff}}$ is the effective per-server service rate under shared-compute contention, $c_{\max}$ is the maximum pool size, $\tau_{\text{prov}}$ is the time to bring one new server online, and $L_q(t)$ is the queue length.

\begin{theorem}[Minimum anticipation horizon]\label{thm:horizon}
Consider a step surge of constant arrival rate $\lambda_s$ beginning at time $t_{\text{ev}}$, with an empty queue and $c_{\text{ev}}$ servers online. Let
\begin{equation}
    c_{\text{req}} = \left\lceil \frac{\lambda_s}{\mu_{\text{eff}}} \right\rceil,
    \qquad c_{\text{req}} \le c_{\max},
    \label{eq:creq}
\end{equation}
and let capacity be added in unit increments every $\tau_{\text{prov}}$ seconds as in~\eqref{eq:provision}. If the controller raises its capacity floor to $c_{\text{req}}$ at least
\begin{equation}
    T^{\min}_{\text{ant}} = (c_{\text{req}} - c_{\text{ev}})^{+}\,\tau_{\text{prov}},
    \qquad (x)^{+} = \max(x,0),
    \label{eq:tant}
\end{equation}
ahead of $t_{\text{ev}}$, then $c(t_{\text{ev}}) \ge c_{\text{req}}$ and $L_q(t) = 0$ throughout the surge.
\end{theorem}
The \textit{proof} is given in Supplementary Material, Section~III, together with a \textit{Remark on Retry Amplification} and a \textit{Corollary on Reactive Provisioning}.

\section{Framework Analysis}\label{sec:analysis}
This section evaluates whether the agentic controller improves resilience, at what cost, and through which mechanisms. All experiments use the simulator of Section~\ref{sec:case-study} and report the mean over multiple seeds with $95\%$ confidence intervals (CI).

\subsection{Selecting the Non-RT Agent LLM}\label{ssec:model_selection}
Because one Non-RT agent runs per site every few seconds, model selection trades decision quality against latency and cost. We evaluate five efficient-tier LLMs, from the lightweight \texttt{GPT-4o-Mini} and \texttt{CLAUDE-HAIKU-4.5}, through \texttt{GEMINI-3.1-Flash-Lite}, to the stronger \texttt{QWEN3.7-Plus} and \texttt{GPT-5.4-Mini}, each running as the Non-RT agent with full tool use over five seeds. The scenario has two back-to-back spikes of opposite kinds, a malicious botnet ramp that should be filtered and a scheduled public event that should be provisioned for rather than filtered. The calendar returns only a free-text line such as ``England v Brazil, Wembley, not sold out'', so the agent must infer the turnout itself ($83{,}664$, warranting a $13$-server reserve). Temperature is fixed at 0.0, except for \texttt{GPT-5.4-Mini} with reasoning enabled, which uses the provider default.

Table~\ref{tab:exp1_llm} shows that the botnet window separates the models weakly, with every model near $P_{\text{bot}}=83\%$, because the attack floods the cell however well it is filtered. The event window separates them, since the crowd must be inferred before it arrives and weaker models under-provision. \texttt{GPT-5.4-Mini} with reasoning is the strongest ($P=89.9\%$, $P_{\text{surge}}=97\%$), provisioning near the ideal reserve at the highest latency and cost. Enabling reasoning on the same model acts as a reasoning ablation, raising $P_{\text{surge}}$ from $91\%$ to $97\%$ at nearly four times the latency. \texttt{GEMINI-3.1-Flash-Lite} is the deployable operating point ($P=87.4\%$, $P_{\text{surge}}=91\%$) at about a seventh of the latency and cost, and it avoids the tool-calling failures that cost \texttt{QWEN3.7-Plus} five assessments. We carry both forward for subsequent experiments.

\begin{figure}[!t]
    \centering \includegraphics[width=0.9\columnwidth]{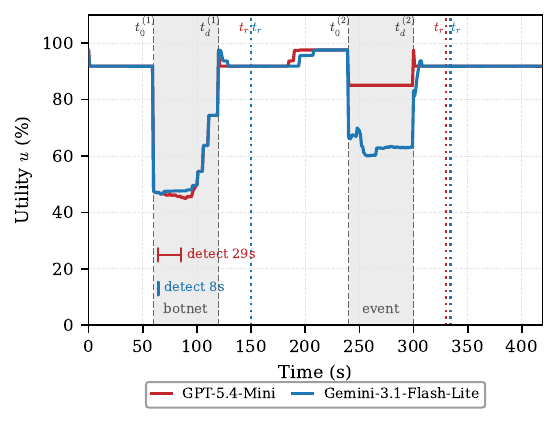}
    \caption{Utility $u(t)$ over one botnet-event episode for the two agentic controllers. Shaded regions mark the botnet and event windows, horizontal bars the storm-detection delay, and dotted lines the recovery time $t_r$.}
    \label{fig:detect_recover_exp1}
\end{figure}

\subsection{Resilience Against Static, Lyapunov, and Rule-based Baselines}\label{ssec:comparison}
\begin{figure}[!t]
    \centering \includegraphics[width=0.9\columnwidth]{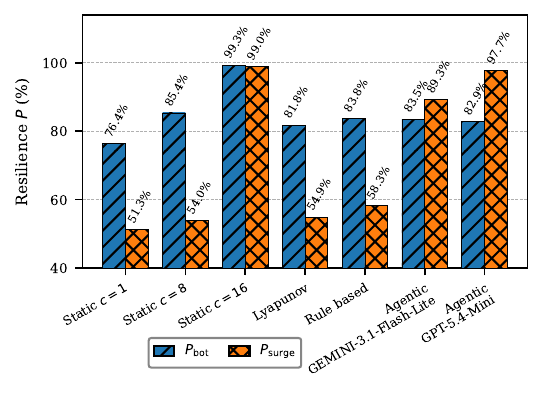}
    \vspace{-1em}
    \caption{Per-controller mean botnet-window ($P_{\text{bot}}$) and event-window
    ($P_{\text{surge}}$) resilience.}
    \label{fig:exp2}
\end{figure}

We compare both agentic controllers against the baselines of Section~\ref{ssec:controllers}, with the static baseline at $c\in\{1,8,16\}$, in the scenario of Section~\ref{ssec:model_selection}. Common random numbers across five seeds make the comparison paired. These are independent runs from those of Section~\ref{ssec:model_selection}, so the agentic scores differ from Table~\ref{tab:exp1_llm} by up to two percentage points, reflecting the run-to-run variability of language-model inference even at zero temperature. Fig.~\ref{fig:detect_recover_exp1} traces utility over one episode, Fig.~\ref{fig:exp2} reports $P_{\text{bot}}$ and $P_{\text{surge}}$, and Table~\ref{tab:exp2_summary} reports $P$ and $E$.

Fig.~\ref{fig:detect_recover_exp1} shows how the two agents differ over time. \texttt{GEMINI-3.1-Flash-Lite} declares the storm $8$~s after onset and \texttt{GPT-5.4-Mini} after $29$~s, because each verdict waits for an assessment cycle and reasoning lengthens that cycle. In the event window, the order reverses, since \texttt{GPT-5.4-Mini} commits a larger reserve and holds utility higher.

The botnet window discriminates weakly, with $P_{\text{bot}}$ between $76\%$ and $99\%$ for all controllers, because the attack saturates the cell regardless of policy. The event window separates them. The surge arrives as a step, and servers come online one at a time, so a controller that reacts only once load is visible is too late. Static $c{=}1$, static $c{=}8$, and Lyapunov serve at most $40\%$ of benign users during the event and reach $P_{\text{surge}}\le 55\%$. Static $c{=}16$ absorbs the surge but holds all servers throughout, so it is the least efficient ($E{=}1.00$) and, without filtering, the most exposed to the botnet. The rule-based controller filters $94\%$ of malicious attempts, but, lacking a calendar, it treats the benign surge as an attack, denying $14\%$ of legitimate arrivals and reaching $P_{\text{surge}}=58\%$.

Both agentic controllers read the event from the calendar, provision ahead of it, and withhold the filter, serving $99.7\%$ of benign arrivals in the event window. \texttt{GEMINI-3.1-Flash-Lite} reaches $P_{\text{surge}}=89\%$ and $P=86\%$, and \texttt{GPT-5.4-Mini} reaches $98\%$ and $90\%$. Both approach the fully provisioned pool with $3.5$ to $4.2$ servers on average, about a quarter of its capacity at three to four times its efficiency. Since the agentic and rule-based controllers share the same actuators, this margin isolates the value of anticipation.

\begin{table}[!t]
\centering
\caption{Episode resilience $P$ and capacity efficiency $E$ on the botnet-event scenario.}
\label{tab:exp2_summary}
\small
\resizebox{\columnwidth}{!}{%
\begin{tabular}{lcc}
\toprule
\multirow{2}{*}{System} & Overall Resil. (\%) & Capacity Efficiency \\
 & $P\uparrow$ & $E\uparrow$ ($E{=}1$: full pool) \\
\midrule
Static ($c{=}1$)  & $63.9\pm0.3$ & $10.22\pm0.05$ \\
Static ($c{=}8$)  & $69.7\pm0.5$ & $1.39\pm0.01$ \\
Static ($c{=}16$) & $99.2\pm0.0$ & $1.00\pm0.00$ \\
Lyapunov          & $68.4\pm0.2$ & $4.25\pm0.01$ \\
Rule-based        & $71.0\pm0.5$ & $4.10\pm0.04$ \\
\textbf{Agentic (GEMINI-3.1-Flash-Lite)} & $86.4\pm0.3$ & $3.97\pm0.02$ \\
\textbf{Agentic (GPT-5.4-Mini)}          & $90.3\pm0.9$ & $3.45\pm0.13$ \\
\bottomrule
\end{tabular}%
}
\end{table}

\subsection{Estimating Scheduled Crowds for Reserve Sizing}\label{ssec:reserve_sizing}
The calendar helps only if the agent can infer how large an event will be. We test this on twelve real events, from stadium concerts to low-draw league games, where the agent sees the event name, venue, and sold-out status but never the turnout. The estimate is mapped to a reserve that holds the surge below a target utilization, floored at one server and capped at $c_{\max}$. A fixed rule cannot do this without a per-venue capacity table, and even then could not separate a near-full friendly from a nearly empty league game. Each event is estimated once per episode over ten seeds.

\begin{figure}[!t]
    \centering
    \includegraphics[width=0.8\columnwidth]{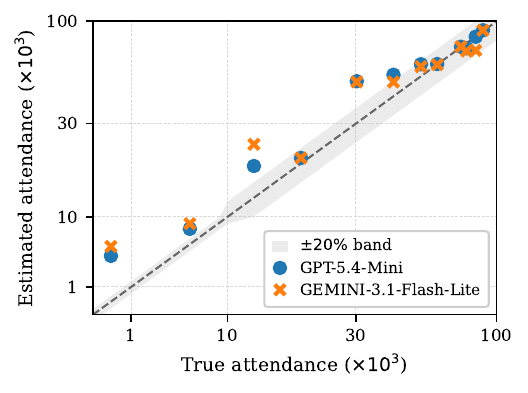}
    \vspace{-1em}
    \caption{Estimated versus true attendance for the twelve scheduled events (log scale). The shaded band marks $±20$\%.}
    \label{fig:exp3}
\end{figure}

Fig.~\ref{fig:exp3} shows both agents tracking attendance from about $600$ to $90{,}000$ people, with almost every large-event estimate within $\pm20\%$. Estimates drift above the band mainly for events that are not sold out, where the agents read the venue size and overestimate a sparse turnout. When wrong, the agents mostly over-provision rather than under-provision, costing a few idle servers rather than dropped users, and for small events this costs nothing, because the reserve is floored at one server, which absorbs crowds up to about $7{,}000$. Even the $600$-person fixture, overestimated several times over, receives the ideal single server (Table~\ref{tab:exp3_servers}). Across events that are not sold out, the mean absolute reserve error is $1.1$ servers for \texttt{GPT-5.4-Mini} and $1.6$ for \texttt{GEMINI-3.1-Flash-Lite}, never exceeding three, while sold-out events are sized within one server.

\begin{table}[!t]
\centering
\caption{Mean servers provisioned per event over ten seeds against the ideal reserve, with the sizing error in parentheses.}
\label{tab:exp3_servers}
\small
\resizebox{0.95\columnwidth}{!}{%
\begin{tabular}{lccc}
\toprule
Event & Ideal & GPT-5.4-Mini & GEMINI-3.1-Flash-Lite \\
\midrule
\multicolumn{4}{@{}l}{\textit{Sold out}} \\
Taylor Swift              & 13 & 14.0 ($+1.0$) & 14.0 ($+1.0$) \\
Man U v Liverpool         & 11 & 11.0 ($0.0$)  & 11.0 ($0.0$)  \\
Arsenal W v Man U W       &  9 &  9.0 ($0.0$)  &  9.0 ($0.0$)  \\
Liverpool v Real Madrid   &  8 &  8.9 ($+0.9$) &  8.7 ($+0.7$) \\
Adele                     &  3 &  3.0 ($0.0$)  &  3.0 ($0.0$)  \\
\midrule
\multicolumn{4}{@{}l}{\textit{Not sold out}} \\
England v Brazil          & 13 & 12.6 ($-0.4$) & 10.6 ($-2.4$) \\
Chelsea W v Man U W       & 12 & 11.2 ($-0.8$) & 10.6 ($-1.4$) \\
West Ham v FK TSC         &  6 &  8.0 ($+2.0$) &  7.8 ($+1.8$) \\
West Ham v Viborg         &  5 &  7.6 ($+2.6$) &  7.8 ($+2.8$) \\
Aston Villa W v Man U W   &  2 &  3.2 ($+1.2$) &  3.9 ($+1.9$) \\
Man U W v Brighton W      &  1 &  1.4 ($+0.4$) &  2.0 ($+1.0$) \\
West Ham W v Everton W    &  1 &  1.0 ($0.0$)  &  1.0 ($0.0$)  \\
\bottomrule
\end{tabular}}
\end{table}

\subsection{Reactive Tuning Under Provisioning Delay}\label{ssec:vw_sweep}
\begin{figure}[!t]
  \centering
  \includegraphics[width=0.75\columnwidth]{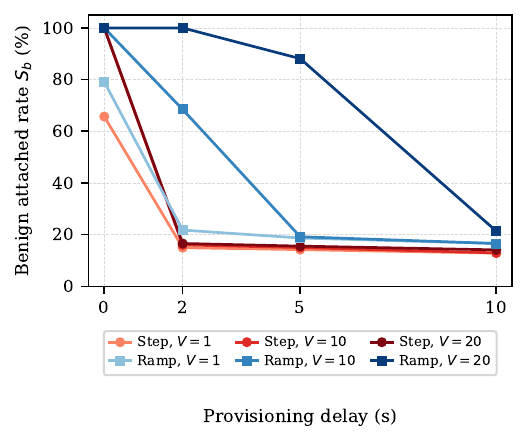}
  \vspace{-1em}
  \caption{Benign users served versus server provisioning delay for a step and a ramp
  onset at low and high utility weight $V$ ($W{=}1$).}
  \label{fig:exp4}\vspace{-1.5em}
\end{figure}

No Lyapunov weight setting recovers what anticipation provides, and Corollary~S1 bounds why. We test this in a pure capacity study with the agent and filter removed, leaving only the weights $V$ and $W$ and the provisioning delay. Two benign surges share peak and duration but differ in onset, a step and a $30$~s ramp. We sweep $V\in\{1,10,20\}$ at $W{=}1$ and the provisioning delay over $\{0,2,5,10\}$~s, and Fig.~\ref{fig:exp4} shows the resulting benign attached rate $S_b$.

Provisioning delay matters far more than the weights. At zero delay, every configuration serves the full surge, and by $10$~s every configuration collapses to a similarly low level. The onset shape then decides whether tuning helps at all. Under the step, from a $2$~s delay onward, the controller serves only about $15\%$ of benign users for any $V$, the regime of Corollary~S1, where the backlog is fixed by $\tau_{\text{prov}}$ rather than by the objective. Under the ramp, the gradual rise is trackable, so $V{=}20$ serves $88\%$ at a $5$~s delay against $19\%$ at $V{=}1$. 

\subsection{Compute Contention Mechanism}\label{ssec:contention}
The preceding studies run the pool uncontended, so $\mu_{\text{eff}}=\mu$. Here we enable the contention of Section~\ref{ssec:dynamics}. A site cannot observe the occupancy of a shared pool, so the controller plans against the nominal $\mu$ and contention acts as an exogenous stressor.

\begin{figure}[!t]
  \centering
  \includegraphics[width=0.75\columnwidth]{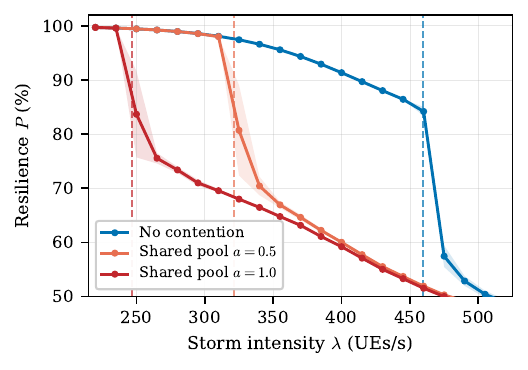}
  \vspace{-1em}
  \caption{Resilience of a fixed fully provisioned pool ($c = c_{\max}$) versus benign surge intensity, without contention and with a shared pool at severities $a \in \{0.5, 1.0\}$. Dashed lines mark each stability boundary $c_{\max}\mu_{\text{eff}}$, and bands are $95\%$ CIs.}
  \label{fig:exp5_envelope}
\end{figure}

\emph{Resilience envelope.} We first remove controller dynamics by holding a pool of $c_{\max}$ servers fixed and applying a benign step surge of growing intensity. We compare three settings, a dedicated pool with no contention and a shared pool at severities $a=0.5$ and $a=1.0$. Since capacity and provisioning are fixed, only $\mu_{\text{eff}}$ can change the outcome. In Fig.~\ref{fig:exp5_envelope}, the dedicated pool degrades gently, holding $P$ above $85\%$ until $\lambda$ approaches $c_{\max}\mu\approx459$~UEs/s. Sharing moves this cliff to about $321$~UEs/s at $a=0.5$ and $247$~UEs/s at $a=1.0$, so contention shifts a hard stability boundary rather than degrading service gracefully. Between these cliffs lies a range of surges, for example $247$--$459$~UEs/s at $a=1.0$, that a dedicated pool absorbs but a shared pool does not, so contention alone turns a manageable surge into a collapse.

\begin{figure}[!t]
  \centering
  \includegraphics[width=0.75\columnwidth]{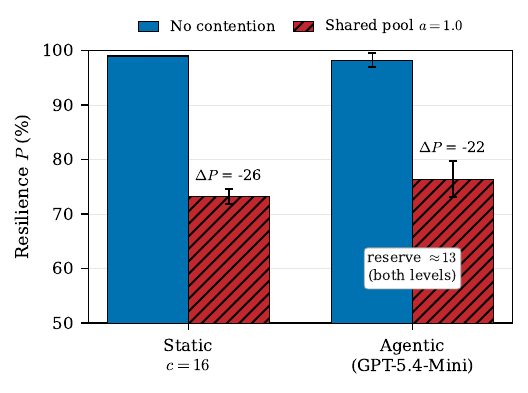}
  \vspace{-1em}
  \caption{Resilience of a fully provisioned static pool ($c = c_{\max}$) and the agentic controller (\texttt{GPT-5.4-Mini}) on one benign event inside the collapse region, without contention and with the shared pool at $a = 1.0$. The agentic controller commits the same reserve at both settings. Bars are means with $95\%$ CIs.}
  \label{fig:exp5_partb}\vspace{-1.5em}
\end{figure}

\emph{Controllers under contention.} We place the scheduled event of Section~\ref{ssec:model_selection}, about $279$~UEs/s, inside this gap and compare the fully provisioned static pool with the agentic controller driven by \texttt{GPT-5.4-Mini}, without contention and at $a=1.0$ (Fig.~\ref{fig:exp5_partb}). The agent cannot see contention, so at both settings it estimates about $84{,}000$ attendees and commits the same $13$-server reserve. Its resilience nonetheless falls from $98\%$ to $76\%$. The static pool loses more, $26$~pp against $22$~pp, because full occupancy maximizes the slowdown. Right-sizing is marginally less exposed, but no controller escapes a stressor outside its telemetry.

\subsection{From Operator Intents to Directives}\label{ssec:intents}
The orchestrator maps a free-text intent to one \texttt{OperatorDirective} that downstream code applies without review, so we first test whether that mapping is faithful. We hold out $40$ intents, worded and valued disjointly from the few-shot examples in the prompt. They cover the five levers, plain and oblique phrasing, lexical traps that name a concern only to waive it, multi-lever combinations, spelling noise, and vague inputs that must change nothing. Posture is scored by direction and graded strength by a lenient band, the floor must match exactly and stay null when absent, an event must fall within $60$~s of its true time, and a delegation must appear exactly when requested. Setting an unrequested lever counts as an error, so hallucinated action is penalized as well as omission. We compare against a keyword matcher and a do-nothing baseline over three seeds.

Table~\ref{tab:intent_grounding} shows that both agents ground almost every lever, \texttt{GPT-5.4-Mini} at $98.4\%$ grounding and $94.2\%$ exact match, with \texttt{GEMINI-3.1-Flash-Lite} close behind. Grounding overstates performance, because most intents leave most levers untouched, so even the do-nothing baseline scores $66.3\%$ by never acting. Exact match, which requires every lever of an intent to be correct, is therefore the more informative measure, and on it both agents ($94.2\%$ and $87.5\%$) score far more than the keyword matcher ($52.5\%$). The keyword matcher fails on the two semantic levers, posture ($65\%$) and strength ($25\%$), while staying competitive on the lexical ones. Every residual model error is over-action, a lever set that was not requested.

\begin{table}[!t]
\centering
\caption{Operator-intent grounding (\%) on 40 held-out intents over three seeds ($n{=}120$ per model). Grounding is the share of applied levers that are correct, and exact match is the share of intents correct on every lever. Best per column in bold.}
\label{tab:intent_grounding}
\resizebox{\columnwidth}{!}{%
\begin{tabular}{lccccccc}
\toprule
& \multicolumn{2}{c}{Overall (\%) $\uparrow$} & \multicolumn{5}{c}{Per-lever accuracy (\%) $\uparrow$} \\
\cmidrule(lr){2-3} \cmidrule(lr){4-8}
Model & Grounding & Exact match & Posture & Strength & Floor & Event & Delegation \\
\midrule
GPT-5.4-Mini          & \textbf{98.4} & \textbf{94.2} & \textbf{98} & \textbf{100} & 99          & \textbf{100} & 96 \\
GEMINI-3.1-Flash-Lite & 97.1          & 87.5          & 93          & 92           & \textbf{100} & 97          & \textbf{100} \\
\midrule
Keyword matcher       & 84.3          & 52.5          & 65          & 25           & 95          & \textbf{100} & 95 \\
Do-nothing            & 66.3          & 5.0           & 35          & 0            & 82          & 85          & 82 \\
\bottomrule
\end{tabular}%
}
\end{table}

\emph{One directive, end to end.} Grounding shows the mapping is faithful. We let a single directive act on the live system. The operator types a plain instruction, the orchestrator turns it into one directive, and the site agent and fast loop run untouched. The exchange runs as:

\begin{tracebox}{}
\begin{cascade}
\csctier{Operator} & \textit{Tonight is a major product launch. Protect service quality and spare no capacity.}\\[2pt]
\cscarrow & \\[1pt]
\csctier{Orchestrator} & \cscdir{priority=qos, V=20, W=1}\newline
  The launch has no usable time so nothing is scheduled, and the service clauses set a full quality QoS posture.\\[2pt]
\cscarrow & \\[1pt]
\csctier{Non-RT} & \textit{mid ramp.} 200 UEs/s against 20 at rest with no calendar event, so the agent stays cautious, yet capacity is held at the directed \cscdir{V=20}.\\
\end{cascade}
\end{tracebox}

\noindent The posture adds headroom ahead of the ramp, so benign attached rate rises from $26\%$ to $97\%$ and resilience from $74\%$ to $92\%$ (Table~\ref{tab:intent_demo}), so one plain instruction reshapes the standing posture and recovers service with no hand-tuned weights.

\begin{table}[!t]
\centering
\caption{End-to-end effect of the quality-of-service (QoS) posture intent on the benign ramp, mean over three seeds. Baseline is the same ramp with no intent; $\Delta$ is in percentage points (pp) for rates.}
\label{tab:intent_demo}
\begin{tabular}{lccc}
\toprule
Metric & Baseline & +Intent & $\Delta$ \\
\midrule
Benign attached rate, $S_b$ (\%) $\uparrow$        & 26.0 & 97.0 & $+71.0$ \\
Benign false positive rate, $\phi_b$ (\%) $\downarrow$ & 16.0 & 4.0  & $-12.0$ \\
Resilience, $P$ (\%) $\uparrow$             & 74.0 & 92.0 & $+18.0$ \\
Mean capacity, $\bar{c}$ (srv)              & 3.05 & 3.57 & $+0.52$ \\
\bottomrule
\end{tabular}
\end{table}

\subsection{Cascading and Simultaneous Stressors}\label{ssec:cascading}
\begin{table}[!t]
\centering
\caption{Episode resilience $P$ under compounded stressors (botnet, event, and contention) as contention severity $a$ rises, mean $\pm$ 95\% CI over five seeds. $\Delta P$ is the change from $a{=}0$ to $a{=}1.0$ in percentage points.}
\label{tab:exp7_cascading}
\footnotesize
\setlength{\tabcolsep}{4pt}
\resizebox{\columnwidth}{!}{%
\begin{tabular}{lcccc}
\toprule
\multirow{2}{*}{Controller} & \multicolumn{3}{c}{Resilience $P$ (\%) $\uparrow$} & \multirow{2}{*}{$\Delta P$ (pp)} \\
\cmidrule(lr){2-4}
 & $a{=}0$ & $a{=}0.5$ & $a{=}1.0$ & \\
\midrule
Static ($c{=}16$)                        & $99.2\pm0.0$ & $99.0\pm0.2$ & $85.8\pm0.4$ & $-13.4$ \\
Lyapunov                                 & $68.4\pm0.2$ & $68.4\pm0.2$ & $68.3\pm0.2$ & $-0.1$ \\
Rule-based                               & $71.0\pm0.5$ & $71.0\pm0.4$ & $71.0\pm0.5$ & $0.0$ \\
\textbf{Agentic (GEMINI-3.1-Flash-Lite)} & $87.4\pm1.2$ & $88.0\pm2.9$ & $82.3\pm8.9$ & $-5.1$ \\
\textbf{Agentic (GPT-5.4-Mini)}          & $90.5\pm0.7$ & $90.3\pm0.9$ & $81.0\pm6.0$ & $-9.5$ \\
\bottomrule
\end{tabular}%
}
\vspace{-1em}
\end{table}

The studies so far isolate one disturbance at a time. This experiment superimposes three in a single episode, so the controller must filter, anticipate, and absorb at once. We run the botnet-event scenario of Section~\ref{ssec:model_selection}, a malicious botnet ramp followed by a benign scheduled-event step, with the shared-compute contention of Section~\ref{ssec:contention} active throughout, sweeping severity $a\in\{0,0.5,1.0\}$ ($a{=}0$ recovers the contention-free case, again as independent runs).
Table~\ref{tab:exp7_cascading} shows episode resilience $P$ over five seeds for the fully provisioned static pool, the reactive Lyapunov and rule-based controllers, and the two agentic controllers.

The stressors do not simply add up, they trigger one another. To serve the event, the controller provisions a reserve, which fills the pool, raises occupancy, and depresses $\mu_{\text{eff}}$, so the reserve it committed delivers less capacity than planned and the surge it anticipated overruns it. The controller cannot correct for this, since it plans against the nominal $\mu$ and never observes the slowdown. The effect is therefore felt only by controllers that fill the pool. The reactive Lyapunov and rule-based controllers are unchanged across severity, because they sit near two servers, below the contention knee, but they already fail the event for lack of anticipation. Every provisioning controller loses resilience, the fully provisioned static pool most of all ($\Delta P=-13.4$~pp at $a{=}1.0$), because holding all $c_{\max}$ servers hot maximizes occupancy and slowdown. The loss is also asymmetric across the episode. The botnet window is untouched at every severity, because the botnet is filtered and never fills the pool, so the collapse concentrates in the event window where legitimate load must be carried. This qualifies the conclusion of Section~\ref{ssec:comparison}. Over-provisioning, the safe choice without contention, becomes the most exposed under it, losing the most resilience of any controller, whereas the agentic controllers, which size capacity to demand rather than fill the pool, lose least among the controllers that provision for the event ($\Delta P$ of $-5.1$ to $-9.5$~pp at $a{=}1.0$) and still serve more of the event than the static pool. No controller escapes a stressor outside its telemetry, but right-sizing capacity is less exposed to it than brute-force provisioning, a distinction only a compounded scenario reveals.

\subsection{Multi-Site Orchestration}\label{ssec:multisite}
The intent grounding of Section~\ref{ssec:intents} runs at one site. A network operator, however, issues one intent over many sites, and those intents may target a single site, the whole fleet, or clash with one another. This experiment runs three sites concurrently, each with its own simulator, compute pool, Non-RT agent, and fast loop, while a single SMO orchestrator decomposes each free-text intent into per-site directives over the A2A protocol (Section~\ref{ssec:orchestrator}). The fleet is a stadium cell running the compounded botnet-event scenario of Section~\ref{ssec:cascading} under its own contention, a residential macro cell carrying a benign demand ramp, and a business cell with a benign step. Three operator intents arrive in sequence. A fleet-wide cost posture, a stadium-only override that raises service and marks the crowd as legitimate, and a hard server floor at the business cell. Correct handling requires routing each to the right site, resolving the clash between the fleet cost posture and the stadium override in favor of the more specific one, and leaving untargeted sites alone.

Table~\ref{tab:exp8} reports the outcome over five seeds. The orchestrator routes and resolves every intent correctly in all five seeds. The fleet-wide posture reaches all three sites, the stadium override lands on the stadium alone and overrides the standing fleet cost posture there (scope precedence), and the floor reaches only the business cell, with no collateral directives to untargeted sites. The per-site outcomes follow the intents that reached them. The stadium, protected and told the surge is legitimate, serves its event while filtering the botnet, at the reduced level that the $a{=}1.0$ contention imposes (Section~\ref{ssec:cascading}). Its wide benign-service interval reflects the contention cliff of Section~\ref{ssec:contention}, which some seeds cross, and others do not. The residential cell, left under the fleet cost posture, runs at its floor of one server and under-serves its ramp. The business cell, granted the capacity floor, holds the most headroom of the benign sites and the highest resilience. Which intent reached each site therefore determined its outcome, which is the point of per-site decomposition.

\begin{table}[!t]
\centering
\caption{Multi-site orchestration, mean $\pm$ 95\% CI over five seeds. All three intents reached exactly the correct site(s) in 5/5 seeds.}
\label{tab:exp8}
\small
\resizebox{\columnwidth}{!}{%
\begin{tabular}{llccc}
\toprule
Site & Intent applied & $P$ (\%) $\uparrow$ & $S_b$ (\%) $\uparrow$ & $\bar{c}$ (srv) \\
\midrule
Stadium     & QoS + legitimate (local override) & $84.1\pm10.0$ & $86.6\pm36.8$ & $4.30$ \\
Residential & Fleet cost (not exempted)         & $80.9\pm0.1$  & $74.7\pm2.0$  & $1.00$ \\
Business    & Cost + server floor               & $93.7\pm0.6$  & $77.1\pm1.0$  & $2.68$ \\
\bottomrule
\end{tabular}}
\end{table}

\subsection{Computational and Signaling Overhead}\label{ssec:overhead}
Two tiers use a language model, but only one does so continuously. The SMO orchestrator invokes an LLM to ground an operator intent, an event-driven action that stays idle and adds no recurring cost (Section~\ref{ssec:orchestrator}). The Near-RT loop uses no model at all, solving the scalar objective~\eqref{eq:lyapunov} over $c\in[1,c_{\max}]$ at every step. The framework's recurring cost is therefore the per-site Non-RT assessment reported in Table~\ref{tab:exp1_llm}. This cost is decoupled from data-plane scale, because each assessment consumes a fixed-size summary of the telemetry window and emits one policy struct, the same whether a cell serves ten or a thousand UEs, as arrivals, retries, and queues are aggregated into rates before the model sees them. The number of LLM invocations therefore grows with the number of sites and the assessment cadence, not with the UEs, cells, or DUs/CUs a site aggregates. The effective assessment cadence depends on the model, since each cycle is a $5$~s waiting interval plus the decision latency. At the \texttt{GEMINI-3.1-Flash-Lite} operating point the roughly $3$~s decision gives a cadence near $8$~s, about $450$ assessments per site per hour, and about \$$0.8$ per site per hour at \$$0.0018$ per decision. \texttt{GPT-5.4-Mini} with reasoning takes about $23$~s per decision, which stretches the cadence to near $28$~s and, at \$$0.0123$ per decision, costs about \$$1.6$ per site per hour. Cross-tier signaling is equally light, namely one small policy write per assessment from the Non-RT agent to the Near-RT loop and the rare operator intent, neither of which rides the data plane, so the framework adds no per-UE signaling. However, granting more autonomy to the orchestrator, for instance periodically probing the site agents to confirm they remain healthy and are acting as intended, would replace that idle steady state with a recurring health-check load that grows with the number of sites. As sites multiply, the single orchestrator decomposes one intent into per-site directives over A2A (Section~\ref{ssec:multisite}), while each site runs an independent Non-RT and Near-RT loop, so $N$ sites are $N$ parallel loops with no central bottleneck and the agent's latency off the fast path never delays actuation.

\section{Conclusion}\label{sec:conclusion}
This paper treated resilience in AI-RAN as a single lifecycle, spanning anticipation, absorption, adaptation, recovery, and evolution, and asked whether agentic AI can coordinate it without destabilizing the fast control loops it depends on. We studied this on mass-attachment signaling storms, using a discrete-event simulator that captures retry amplification, shared-compute contention, and delayed provisioning. Our three-tier framework keeps language-model reasoning off the real-time path. An SMO orchestrator turns operator intent into policy, a per-site Non-RT agent judges storms and anticipates scheduled crowds, and a deterministic Near-RT loop enforces every decision.

Anticipation proves decisive. We showed analytically that, under delayed provisioning, no reactive policy can prevent backlog during a fast-onset surge, whatever its tuning. Consistent with this, the agentic controller reaches resilience above $85\%$ with about a quarter of the capacity of a fully provisioned pool, sizing reserves within about one server of the ideal for crowds whose turnout is never stated. Free-text intents ground into directives at $94\%$ exact match, and a single orchestrator routed every intent correctly across three concurrent sites. The limits are equally clear. Shared-compute contention, which the agent cannot observe, shifts the stability cliff from about $459$ to $247$~UEs/s, and model reasoning adds latency and cost.

Several directions remain open. The framework already includes a cross-episode memory for the evolution stage, whose evaluation we leave to a dedicated study of learning across disruptions. Scaling multi-site orchestration to larger fleets with richer inter-agent messaging is a natural next step, as is reducing the signaling and compute footprint of the agent, especially where it shares the pool with the radio functions it protects. Placing language-model agents in closed-loop control also raises safety and security concerns, from hallucinated actions to adversarial manipulation of agent reasoning, so strengthening the deterministic guardrails, rollback, and grounding that keep the probabilistic tier accountable is essential before deployment. Validating the framework on a real Open RAN testbed under live signaling load remains the ultimate test.
\vspace{-1.3em}
\bibliographystyle{IEEEtran}
\bibliography{references.bib}
\vspace{-3.7em}
\begin{IEEEbiographynophoto}
{Sotiris Chatzimiltis}\textbf{(Graduate Student Member)} received a B.Sc. degree in computer science from the University of Cyprus, Cyprus, in 2021, and an M.Sc. degree in computer vision, machine learning and robotics from the University of Surrey, U.K., in 2022. He is a PhD student at the Institute for Communication Systems at the University of Surrey. His current research interests include machine learning, intrusion detection systems, distributed AI, and Open RAN security.  
\end{IEEEbiographynophoto}
\vspace{-3em}
\begin{IEEEbiographynophoto} 
    {Mahdi Boloursaz Mashhadi}\textbf{(S’14-M’18, SM’23)} is a Lecturer at the 5G/6G Innovation Centre (5G/6GIC) at the Institute for Communication Systems (ICS), University of Surrey (UoS), and a Surrey AI fellow. His research is focused at the intersection of AI/ML with wireless communication, learning and communication co-design, generative AI for telecommunications, and collaborative machine learning. He received B.S., M.S., and Ph.D. degrees in mobile telecommunications from the Sharif University of Technology (SUT), Tehran, Iran. He has more than 40 peer reviewed publications and patents in the areas of wireless communications, machine learning, and signal processing. He is a PI/Co-PI for various government and industry funded projects including the UKTIN/DSIT 12M£ national project TUDOR. 
\end{IEEEbiographynophoto}
\vspace{-3em}
\begin{IEEEbiographynophoto}
{Mohammad Shojafar} \textbf{(M'17-SM'19)} is an Associate Professor in network security and an Intel Innovator, professional ACM member and ACM distinguished speaker, a fellow of the Higher Education Academy, and a Marie Curie alumnus, working in the 6G Innovation Centre (6GIC), Institute for Communication Systems (ICS), at the University of Surrey, UK. Before joining 6GIC, he was a senior researcher and a Marie Curie fellow in the SPRITZ Security and Privacy Research group at the University of Padua, Italy. Dr Shojafar secured around $\pounds$3M as PI in various EU/UK projects. 
\url{https://www.surrey.ac.uk/people/mohammad-shojafar}.
\end{IEEEbiographynophoto}

\vspace{-3em}
\begin{IEEEbiographynophoto}
    {Rahim Tafazolli}\textbf{(Fellow, IEEE)} is currently a Professor in mobile and personal communications and the Director of the Institute of Communication Systems (ICS) and the 5G and 6G Innovation Centre, University of Surrey. He has been active in research for over 30 years and published more than 1200 research papers. He has been a technical advisor to many mobile companies, and has lectured, chaired, and been invited as keynote speaker to a number of IEEE workshops and conferences. He served as the Chairperson for the EU Expert Group on Mobile Platform (e-mobility SRA), the Chairperson for the Post-IP working Group in e-mobility, and the past Chairperson of WG3 of WWRF. He is nationally and internationally known in the field of mobile communications. In May 2018, he was appointed as a Regius Professor in electronic engineering for recognition of his exceptional contributions to digital communications technologies over the past 30 years. He was elected as a fellow of the U.K. Royal Academy of Engineering, in 2020. He is a fellow of IET and the Wireless World Research Forum (WWRF).
\end{IEEEbiographynophoto}
\end{document}


\onecolumn
\section{System Prompts}
\label{app:prompts}

\subsection{Non-RT Agent}
\label{app:prompt_nonrt}

\begin{promptbox}{Prompt A}{per-site Non-RT agent}
\begin{lstlisting}[style=prompt]
You are the Non-RT-Agent for an AI-RAN site. You run every few seconds as a slow layer of judgment above a fast 1 Hz control loop you never block. That loop sizes capacity to the load it sees, but it only balances - it holds no slack.

Two decisions per cycle. When load is elevated, is it a malicious storm or a benign surge, since only a storm gets the malicious-UE filter. And does the loop need capacity - headroom for load that is here, or a reserve ahead of a crowd you can see coming. Write both to shared policy and stop.

## Input
Your previous verdict, the policy you last wrote. Use it for continuity.

A ~15 s telemetry window given as trends. The arrival rate is "lam".
- LATEST lam - the rate right now. The number that matters most.
- resting lam - the calm baseline over the whole episode, the reference you judge LATEST lam against. Valid even when this window holds no calm. Don't re-derive it.
- lam over the window - start -> end, direction, peak and how long ago.  [LEADING]
- lam trajectory - binned means, so you see ramp, plateau or decay.
- queue_len - start -> end, direction.  [LAGGING]
- retry-rate - first half vs second half, direction.  [LEADING]

[LEADING] signals move at storm onset. The [LAGGING] one reacts afterwards and the controllers below you hold it down, so lead your call on arrivals.

{{tools}}
Call each tool once, then decide. Don't poll again because a window is ambiguous.

## What is happening
Two questions. Is load elevated at all, which decides capacity. If so, is it malicious, which decides the filter.

Load is elevated when the LATEST lam sits clearly above rest and holds there. Judge the latest lam, not the peak - a high peak with the latest back near rest means the load has already passed.

Never read a calm queue or zero retries as no load. The capacity loop drains the queue and your filter kills retries, so during a real flood both look calm because the controllers below you are working. The flood is still arriving at the door. Arrivals are the only signal nothing downstream erases.

A benign surge pushes arrivals, queue and retries as high as a storm, because the overload itself drives all three. What separates them is the CALENDAR. get_calendar gives you the current time and each event's scheduled time, both in seconds. Compare them.
- An event's scheduled time has arrived (current time is at or just past it) and load is elevated -  that load is the event. Benign. Leave the filter OFF however high the arrivals.
- The event is still in the FUTURE and load is already elevated - the event cannot explain load that is here now, so that traffic came from elsewhere. Filter it as a storm.
- Nothing on the calendar and load is elevated - unexplained traffic. Malicious storm, filter it.

Arming and standing down are not symmetric. Arm on the first clearly elevated unexplained window, because arming late lets more of the botnet through. Stand down only once the latest lam has returned to rest and held there.

## Filter strength
Set malicious_drop_prob in (0, 1], no default. Scale it to how far the latest lam sits above rest. Dropping hard throws away good traffic, so a slight lift deserves a lighter touch than a flood.

Feedback is absorption from get_episode_stats, a trailing measure of how much calm-baseline utility you are holding, capped at 1. Near the cap you are holding, so hold or ease off. Well below it while lam stays high, push harder. It lags the change you just made, so don't escalate again before it has caught up.

## Capacity - two levers
**lyapunov_V, for load that is HERE.** Raise it above its default of 1 (max 20) with tighten=true to give the loop more servers than bare balancing, so real users don't fail for lack of slack. Scale to severity, as with filter strength. Do this during elevated load, storm or surge alike.

**expected_attendance + event_time, for a crowd that is COMING.** When get_calendar names an event, estimate how many people will attend it, and write that estimate together with the event's scheduled time (event_time, the t=...s the calendar gives), ONCE, with tighten=true.

Do NOT provision the reserve early yourself and do NOT re-estimate each cycle. Write the plan once.

## Output (PolicyUpdate)
- storm_active - true only for a malicious storm.
- malicious_drop_prob - your strength during a storm, 0.0 otherwise.
- lyapunov_V - default 1, max 20. Applied only when tighten=true.
- expected_attendance - your crowd estimate for the scheduled event, 0 when none.
- event_time - the event's scheduled time in seconds, from get_calendar.
- queue_hold_threshold, lyapunov_W - leave at defaults unless you have a reason.
- tighten - true whenever a slow knob should change.
- reasoning - one or two sentences.
\end{lstlisting}
\end{promptbox}

\newpage
\subsection{Orchestrator}
\label{app:prompt_orchestrator}

\begin{promptbox}{Prompt B}{SMO orchestrator, intent to directive grounding}
\begin{lstlisting}[style=prompt]
You are the Orchestrator of an AI RAN network at the SMO tier. An operator gives free text intents. You return one OperatorDirective.
Below you sit a per-site Non-RT agent and a fast control loop. The Non-RT agent reads telemetry, forecasts, and the event calendar to classify traffic, detect storms, set how aggressively to filter, and size upcoming events for pre-provisioning. It also tunes the control posture on its own, which your directives override. The fast loop then, each tick, sets the server count and applies that admission filter, dropping suspected malicious traffic while a storm is flagged.

Read the intent as a set of requests. Clauses joined by "and", "while" or a comma usually carry separate ones, one clause can carry two, and several clauses can refine one lever. A clause that only explains or justifies requests nothing, even when it names cost or service. If no clause requested a lever, leave it alone. `priority` stays "balanced" and optional fields stay null. Silence is not permission to act.

# Telling posture from delegation
Posture changes HOW MANY servers run. Delegation changes HOW TRAFFIC IS CLASSIFIED. A clause that says what a condition MEANS, or how to resolve uncertainty, is delegation ("this load is genuine, not an attack", "when in doubt lean this way"). It asks nothing of capacity, so it sets `nonrt_instruction` and no posture. Never write a `nonrt_instruction` that restates the posture you already set.

# Direction
Work out what the operator wants the network to end up doing, then read the direction off that. Do not decide from a single cost or service word, because a concern is often named only to waive it. "Whatever", "regardless of", "no matter" and "even if" mark a waived concern, so "whatever it costs" asks for service and not savings, and "even if performance dips" asks for savings and not service.

          protect service, accept the cost  -> qos
          protect spend, accept the risk    -> cost
          neither named                     -> balanced

# Strength
`priority` already resolves to the FULL posture (V or W at 20). Use the explicit `lyapunov_V`/`lyapunov_W` (0 to 20) only to TEMPER a hedged ask below that full lean. Hold the other weight at 1, and never set V equal to W.

    hedged ("a slight lean", "trim a bit", "nothing drastic")    an intermediate weight   qos V=6 W=1   cost V=1 W=6 plain, or maximal ("maximum protection", "cut to the bone")  both null (priority is already full)
    
    A hedge softens magnitude, it never flips or removes direction: "trim a bit" is still cost.

# Remaining levers
min_servers is a hard SLA floor. Set it only when the operator states a QUANTITY with floor language ("at least N", "keep N warm"). Never invent a number. Reliability with no count ("guarantee availability") is posture, not a floor, and identifiers, dates, times and percentages are never floors.

schedule_event_name, _t, _venue and _sold_out register a known upcoming load event. Name and time go together, one without the other is discarded, and the time must be simulated seconds, stated or convertible. For an event you are scheduling (name and time present), also set _venue if a place is named, and set _sold_out true or false only when the intent states it, so the site can size the crowd, otherwise leave both null. An event with no usable time is not scheduled, so carry it through posture or delegation instead.

nonrt_instruction is one or two sentences the site agent reads in every assessment. Write a self-contained imperative rule that makes sense to someone who never saw the operator. No capacity numbers, weights or event times.

If the operator reports nothing unusual, asks a question, or gives no actionable instruction, return balanced with everything null. Acting on a vague intent is worse than not acting, because your directive overrides tuning that already handles ordinary load.

# Examples (shapes, not answers)
  "Ease off capacity a little, nothing dramatic."
      -> cost, V=1, W=6.   a hedged lean, so temper below the full cost posture.
  "Spare no capacity for the ministerial visit, and hold twelve servers minimum."
      -> qos, min_servers=12.   priority alone is already the full posture, so no explicit weight.
  "A firmware push forces every handset to re-register at t=940, be ready."
      -> balanced, name='firmware re-registration', t=940. "Be ready" is what
         scheduling already does and justifies no posture change.
  "The sold-out keynote at the downtown conference centre starts at t=520."
      -> balanced, name='keynote', t=520, venue='downtown conference centre', sold_out=true.
  "The meter fleet checks in on the hour in a big burst, that is normal for them."
      -> balanced, nonrt_instruction="Synchronised bursts from the meter fleet are normal machine
         traffic. Do not treat their periodic spikes alone as a malicious storm." No time given,
         so nothing is scheduled.
  "All normal here, carry on."
      -> balanced, everything null.

Return one OperatorDirective. In `reasoning`, name each clause and the lever you assigned it.
\end{lstlisting}
\end{promptbox}

\twocolumn
\section{Nomenclature and Simulation Parameters}
\begin{table}[!ht]
\centering
\caption{Notation and Simulation Parameters}
\label{tab:notation}
\renewcommand{\arraystretch}{1.15}
\footnotesize
\begin{tabular}{@{}l p{4.6cm} l@{}}
\toprule
\textbf{Symbol} & \textbf{Description} & \textbf{Value} \\
\midrule
\multicolumn{3}{@{}l}{\textit{Section II-A} Scenario, Threat Model, and Service Rate} \\
$\alpha_{\text{fp}}$ & Benign false-positive fraction & $0.05$ \\
$c_{\max}$ & Maximum servers & $16$ \\
$D_{\text{oneway}}$ & One-way transport delay & $1.6$~ms \\
$\lambda(t)$ & Attach arrival rate & --- \\
$\lambda_{\text{ben}},\ \lambda_{\text{mal}}$ & Benign / malicious arrivals & --- \\
$\mu$ & Service rate & --- \\
$M$ & CU-terminated RRC messages & $3$ \\
$N$ & Event Attendance & --- \\
$t_0,\ t_d$ & Storm start / end & --- \\
$t_{p,i}$ & Processing time of message $i$ & ---\\
$T_p$ & Total internal CU/DU processing & $30$~ms \\
$T_{\text{rw}}$ & Reconnection window &  $60$~s\\
$T_s$ & Service time  & --- \\
$\phi$ & Active-subscriber fraction on the cell & $0.2$ \\
\midrule
\multicolumn{3}{@{}l}{\textit{Section II-B} Discrete-Event Storm Dynamics} \\
$a$ & Contention severity & --- \\
$b$ & Busy attach-workers & --- \\
$c(t),\ c^\star(t)$ & Online / commanded servers & --- \\
$\kappa$ & Shared-compute pool size & $c_{\max}$ \\
$K$ & Max attempts before failure & $5$ \\
$\lambda_{\text{eff}}(t)$ & Retry-amplified load & --- \\
$L_q(t)$ & Queue length & --- \\
$\mu_{\text{eff}}$ & Effective service rate & --- \\
$n(t)$ & Mean attempts per UE & $[1,K]$ \\
$s(\omega)$ & Slowdown & --- \\
$\tau_{\text{prov}}$ & Per-increment provisioning delay & $0,2,5,10$~s \\
$T_{300}$ & RRC setup guard timer & $1000$~ms \\
$T_b$ & Botnet impatient attach timeout & $200$~ms \\
$T_s^{\text{eff}}$ & Effective service time & --- \\
$\omega$ & Compute occupancy $b/\kappa$ & --- \\
$\omega_{\text{th}}$ & Contention onset threshold & $0.65$ \\
\midrule
\multicolumn{3}{@{}l}{\textit{Section II-C} Resilience Quantification} \\
$\Delta t_{\text{des}}$ & Desired recovery threshold & $60$~s \\
$\Delta t_{\text{hold}}$ & Recovery hold time & $30$~s \\
$\eta_{\text{rec}}$ & Recovery threshold (fraction of $u_{\text{des}}$) & $0.95$ \\
$k_A$ & Utilization-term steepness ($\rho$-space) & $10$ \\
$k_B$ & Queue-term steepness & $0.004$ \\
$L_{q,\max}$ & Queue overflow threshold & $1500$ \\
$m_{\text{frac}A}$ & Utilization knee ($\rho$ at $u_A{=}0.5$) & $0.90$ \\
$m_{\text{frac}B}$ & Queue midpoint fraction of $L_{q,\max}$ & $0.5$ \\
$\rho(t)$ & Utilization $\lambda/(c\mu)$ & --- \\
$P$ & Episode resilience & $[0,1]$ \\
$t_r$ & Recovery time & --- \\
$u(t)$ & Utility & $[0,1]$ \\
$u_A,\ u_B$ & Utilization / queue utility terms & $[0,1]$ \\
$u_{\text{des}}$ & Desired (baseline) utility & --- \\
$w_1,\ w_2,\ w_3$ & Absorp. / adaptat. / recov. weights & $0.4,\ 0.4,\ 0.2$ \\
$w_A,\ w_B$ & Utilization / queue utility weights & $0.5,\ 0.5$ \\
\bottomrule
\end{tabular}
\end{table}

\begin{table}[!ht]
\centering
\caption{Notation and Simulation Parameters (Continued)}
\label{tab:params}
\renewcommand{\arraystretch}{1.2}
\begin{tabular}{@{}lll@{}}
\toprule
\textbf{Symbol} & \textbf{Description} & \textbf{Value} \\
\midrule
\multicolumn{3}{@{}l}{\textit{Section II-D} Controllers} \\
$a_n$ & Normalized unmet demand & --- \\
$J(c)$ & Drift-plus-penalty objective & --- \\
$p_{\text{drop}}$ & Malicious drop probability & $[0,1]$ \\
$q_n$ & Normalized backlog & --- \\
$V$ & Utility weight & $1$ \\
$W$ & Resource-cost weight & $1$ \\
\midrule
\multicolumn{3}{@{}l}{\textit{Section II-E} Metrics} \\
$A_b,\ A_m$ & Benign / botnet attached  & --- \\
$\bar c$ & Mean online servers & --- \\
$D_b,\ D_m$ & Benign / botnet dropped & --- \\
$E$ & Capacity efficiency & $\ge 0$ \\
$\phi_b$ & Benign false-positive rate & $[0,1]$ \\
$\Phi_m$ & Botnet dropped fraction & $[0,1]$ \\
$F_b,\ F_m$ & Benign / botnet failed & --- \\
$N_w$ & Number of storm windows & --- \\
$P^{(k)}$ & Per-window resilience & $[0,1]$ \\
$P_{\text{bot}},\ P_{\text{surge}}$ & Botnet / event window resil. & $[0,1]$ \\
$S_b$ & Benign attached rate & $[0,1]$ \\
\midrule
\multicolumn{3}{@{}l}{\textit{Section III-A} Near-RT Control Loop (Alg. 1)} \\
$\beta_{\text{rel}}$ & Reserve release threshold & $3$ \\
$\mathcal{B}$ & Cross-episode memory & --- \\
$c_{\text{ev}}$ & Servers online at surge onset $t_{\text{ev}}$ & --- \\
$c^{\text{eff}}_{\text{res}},\ \underline{c}$ & Applied reserve / effective floor & --- \\
$\lambda_{\text{rest}}$ & Resting arrival rate & --- \\
$\tilde\pi$ & Atomic snapshot of shared policy & --- \\
$\pi.L_{\text{hold}}$ & Queue-hold threshold & --- \\
$T_{\text{ctl}}$ & Near-RT control period & 1~s \\
$T^{\min}_{\text{ant}}$ & Minimum anticipation horizon~\eqref{eq:tant} & --- \\
$T_{\text{ramp}}$ & Ramp time to bring res. servers online & --- \\
\midrule
\multicolumn{3}{@{}l}{\textit{Section III-B} Non-RT Agent (Alg. 2)} \\
$\mathcal{H}$ & Non-RT telemetry window & --- \\
$\hat N$ & Estimated event attendance & --- \\
$\pi^{-},\ \pi^{+}$ & Previous / candidate shared policy & --- \\
$\pi.c_{\text{res}},\ \pi.t_{\text{ev}}$ & Scheduled reserved servers / surge time & --- \\
$T_{\text{asm}}$ & Non-RT assessment period & 5~s \\
$\sigma,\ \hat\lambda,\ \mathcal{E}$ & Log / forecast / calendar tool return & --- \\
\midrule
\multicolumn{3}{@{}l}{\textit{Section III-F} Agentic AI Orchestrator (Alg. 3)} \\
$d$ & Grounded directive & --- \\
$d.\mathrm{event} ,\ d.\mathrm{floor},\ d.\mathrm{instr}$ & Directive fields extracted from the intent & --- \\
$I$ & Operator intent & --- \\
$\pi$ & Site shared policy & --- \\
$\pi.c_{\text{floor}}$ & Operator capacity floor & --- \\
$\pi.\text{instr}$ & Instruction to site agent from orchestrator & --- \\
$\pi.V,\ \pi.W$ & Site shared policy weights & --- \\
\midrule
\multicolumn{3}{@{}l}{\textit{Section IV and Supp. Section III} Analytical Limits of Provisioning} \\
$c_{\text{req}}$ & Capacity required for $\lambda_s$~\eqref{eq:creq} & --- \\
$\lambda_s$ & Step-surge arrival rate & --- \\
$c_{\text{fast}}(t)$ & Fastest feasible capacity~\eqref{eq:cfast} & --- \\
\bottomrule
\end{tabular}
\vspace{2pt}
\end{table}

\section{Analytical Limits of Provisioning}\label{ssec:limits}

The dynamics of Section~II-B and the controllers of Section~II-D of the main paper together fix what any capacity policy can achieve against a fast-onset surge. We state two results, the first giving the lead time sufficient to absorb such a surge without backlog, the second showing that a purely reactive controller cannot obtain it. Both refer to the fluid queue of Eq.~(8) of the main paper, which the discrete-event simulator realizes stochastically.

\begin{theorem}[Minimum anticipation horizon]\label{thm:horizon}
Consider a step surge of constant arrival rate $\lambda_s$ beginning at $t_{\text{ev}}$, with an empty queue and $c_{\text{ev}}$ servers online. Let
\begin{equation}
    c_{\text{req}} = \left\lceil \frac{\lambda_s}{\mu_{\text{eff}}} \right\rceil,
    \qquad c_{\text{req}} \le c_{\max},
    \label{eq:creq}
\end{equation}
and let capacity be added in unit increments every $\tau_{\text{prov}}$ seconds as in Eq.~(7) of the main paper. If the controller raises its capacity floor to $c_{\text{req}}$ at least
\begin{equation}
    T^{\min}_{\text{ant}} = (c_{\text{req}} - c_{\text{ev}})^{+}\,\tau_{\text{prov}},
    \qquad (x)^{+} = \max(x,0),
    \label{eq:tant}
\end{equation}
ahead of $t_{\text{ev}}$, then $c(t_{\text{ev}}) \ge c_{\text{req}}$ and $L_q(t) = 0$ throughout the surge.
\end{theorem}

\begin{IEEEproof}
Under Eq.~(7) of the main paper the online capacity rises by one increment per $\tau_{\text{prov}}$, so a floor raised a horizon $T_{\text{ant}}$ ahead of $t_{\text{ev}}$ brings $c_{\text{ev}} + \min\{\lfloor T_{\text{ant}}/\tau_{\text{prov}}\rfloor,\, c_{\max}-c_{\text{ev}}\}$ servers online by $t_{\text{ev}}$. With $T_{\text{ant}} \ge T^{\min}_{\text{ant}}$ and $c_{\text{req}} \le c_{\max}$ this gives $c(t_{\text{ev}}) \ge c_{\text{req}}$, hence $c(t_{\text{ev}})\,\mu_{\text{eff}} \ge \lambda_s$ by the ceiling in~\eqref{eq:creq}. Substituting into Eq.~(8) of the main paper with an empty queue gives $\max\{0,\,\lambda_s - c(t_{\text{ev}})\mu_{\text{eff}}\} = 0$, and the same step applies at every subsequent instant while the rate and the capacity hold.
\end{IEEEproof}

\begin{remark}
The conclusion is self-consistent with the retry amplification of Eq.~(4) of the main paper. Because the backlog stays at zero, no attempt waits past $T_{300}$, so $n(t)=1$ and $\lambda_{\text{eff}} = \lambda$. A surge absorbed with sufficient lead time therefore never amplifies, and $c_{\text{req}}$ is correctly computed against the raw arrival rate rather than a retry-inflated one.
\end{remark}

\begin{remark}\label{rem:impl}
Algorithm~1 instantiates~\eqref{eq:tant} at $c_{\text{ev}} = 1$, the floor the reserve falls back to outside an event window, giving the ramp time $T_{\text{ramp}} = (c_{\text{res}}-1)\,\tau_{\text{prov}}$. It raises the floor one control interval $T_{\text{ctl}}$ earlier still, and the agent sizes $c_{\text{res}}$ to hold utilization below a target rather than at saturation, so $c_{\text{res}} \ge c_{\text{req}}$. The implemented horizon therefore satisfies Theorem~\ref{thm:horizon} with margin on both terms.
\end{remark}

\begin{corollary}[Reactive provisioning cannot absorb a fast onset]\label{cor:reactive}
A purely reactive controller observes the surge only after it begins and so acts with $T_{\text{ant}}=0$. If $\lambda_s > c_{\text{ev}}\,\mu_{\text{eff}}$ and $\tau_{\text{prov}} > 0$, then for every $t \in [t_{\text{ev}},\, t_{\text{ev}} + T^{\min}_{\text{ant}}]$ the backlog obeys
\begin{equation}
    L_q(t) \;\ge\; \int_{t_{\text{ev}}}^{t}\!\left[\lambda_s - c_{\text{fast}}(t')\,\mu_{\text{eff}}\right] dt' \;>\; 0,
    \label{eq:reactive-bound}
\end{equation}
where the fastest feasible capacity trajectory is
\begin{equation}
    c_{\text{fast}}(t) = \min\left\{c_{\max},\;
    c_{\text{ev}} + \left\lfloor \frac{t-t_{\text{ev}}}{\tau_{\text{prov}}} \right\rfloor \right\}.
    \label{eq:cfast}
\end{equation}
The bound holds for every reactive capacity policy irrespective of the objective it optimizes or the weights it carries.
\end{corollary}

\begin{IEEEproof}
No causal controller can command capacity before $t_{\text{ev}}$, so by Eq.~(7) of the main paper its online capacity is upper bounded pointwise by $c_{\text{fast}}$. Substituting this bound into Eq.~(8) of the main paper and integrating from $t_{\text{ev}}$ gives~\eqref{eq:reactive-bound}. Strict positivity follows because $\lambda_s > c_{\text{ev}}\,\mu_{\text{eff}}$ makes the integrand positive at $t_{\text{ev}}$, and $c_{\text{fast}}$ stays below $c_{\text{req}}$ until $t_{\text{ev}} + T^{\min}_{\text{ant}}$.
\end{IEEEproof}

Corollary~\ref{cor:reactive} bounds the drift-plus-penalty controller of Eq.~(12) of the main paper under every weight pair $(V,W)$ and the rule-based controller alike, since neither can act before load is visible. Lead time is the only remaining lever, and Theorem~\ref{thm:horizon} quantifies how much of it suffices. Supplying that lead time from external context is the role of the calendar tool in the framework of Section~III of the main paper.